\documentclass[floatfix,amsmath,amssymb,aps,prc,twocolumn,superscriptaddress,showpacs,preprintnumbers,nofootinbib]{revtex4-1}

\usepackage{graphicx,color}
\usepackage{multirow}
\usepackage{times}
\usepackage{bm}
\usepackage{epstopdf}
\usepackage{enumerate}
\usepackage{ulem}
\usepackage{hyperref}
\usepackage{physics}

\newcommand{\comment}[1]{}

\renewcommand\sout{\bgroup \color{red} \ULdepth=-.5ex \ULset}

\graphicspath{{./Figs/}}

\makeatletter
\def\@fnsymbol#1{\ensuremath{\ifcase#1\or \dagger\or \ddagger\or
   \mathsection\or \mathparagraph\or \|\or **\or \dagger\dagger
   \or \ddagger\ddagger \else\@ctrerr\fi}}
    \makeatother
\newcommand{\deceased}[1]{\altaffiliation{#1}}
\DeclareMathOperator{\diag}{diag}
\newcommand{\cm}{\mathrm{cm}}

\begin{document}

\title{
A Poincar\'e covariant cascade method
for high-energy nuclear collisions
}
\author{Yasushi Nara}
\affiliation{
Akita International University, Yuwa, Akita-city 010-1292, Japan}
\author{Asanosuke Jinno}
\affiliation{
Department of Physics, Faculty of Science,
Kyoto University, Kyoto 606-8502, Japan}

\author{Tomoyuki Maruyama}
\affiliation{College of Bioresource Sciences, Nihon University,
Fujisawa 252-0880, Japan}

\author{Koichi Murase}
\affiliation{Yukawa Institute for Theoretical Physics,
Kyoto University, Kyoto 606-8502, Japan}

\author{Akira Ohnishi}
\deceased{Deceased.}
\affiliation{Yukawa Institute for Theoretical Physics,
Kyoto University, Kyoto 606-8502, Japan}

\date{\today}

\preprint{YITP-23-77, KUNS-2969}
\begin{abstract}
We present a Poincar\'e covariant cascade algorithm
based on the constrained Hamiltonian dynamics
in an $8N$-dimensional phase space
to simulate the Boltzmann-type two-body collision term.
We compare this covariant cascade algorithm with traditional $6N$-dimensional phase-space cascade algorithms.
To validate the covariant cascade algorithm, we perform box calculations.
We examine the frame dependence of the algorithm in a one-dimensionally expanding system as well as
the compression stages of colliding two nuclei.
We confirm that our covariant cascade method is reliable
to simulate high-energy nuclear collisions.
Furthermore, we present Lorentz-covariant equations of motion for the $N$-body system interacting
via potentials, which can be efficiently solved numerically.
\end{abstract}

\maketitle

\section{Introduction}

High-energy heavy-ion collisions offer a unique opportunity
to study hot and dense nuclear matter~\cite{luo2022properties,Sorensen:2023zkk}.
To extract equilibrium properties of nuclear matter, such as equations of state,
it is necessary to describe the dynamics of finite and non-equilibrium systems using a theoretical transport approach.
Transport models have been widely used
in the past few decades
to understand collision dynamics and experimental data
in high-energy heavy-ion collisions.
These models include parton cascade
models~\cite{Geiger:1997pf,Zhang:1997ej,Molnar:2000jh,Xu:2004mz,Lin:2004en},
Boltzmann-Uhlenbeck-Uehling (BUU) models~\cite{Bertsch:1988ik},
and quantum molecular dynamics (QMD) approaches~\cite{Aichelin:1991xy},
as well as their relativistic extensions~\cite{Buss:2011mx,Sorge:1989dy,Maruyama:1991bp}.
Extensive comparisons of various transport models can be found
in Refs.~\cite{TMEP:2016tup,TMEP:2017mex,TMEP:2019yci,TMEP:2021ljz,TMEP:2022xjg}.

One of the main ingredients of non-equilibrium microscopic transport approaches is
a Boltzmann-type collision term, which is often simulated by the so-called cascade method, e.g.,
a scattering happens when the closest
distance between two particles is less than the interaction range given by
the cross section $d=\sqrt{\sigma/\pi}$, where $\sigma$ is the total cross
section.
However, this geometrical interpretation of cross section
often leads to a violation of Lorentz covariance
for the $N$ interacting particle system
in the $6N$-dimensional phase-space approach.
To address this issue,
numerical codes employ effective algorithms to approximate the relativistic effect in the modeling.
Despite the lack of covariance in these transport models,
reasonable results are obtained when a simulation is performed in the global center-of-mass frame of colliding two nuclei.
It is important to note that different collision algorithms lead to different numerical
results~\cite{Zhang:1996gb,Cheng:2001dz,TMEP:2017mex,TMEP:2019yci,Zhao:2020yvf}.
The causality problems in the cascade models were discussed in
Refs.~\cite{Kodama:1983yk,Kortemeyer:1995di,Zhang:1996gb,Zhang:1998tj,Xu:2004mz,Zhao:2020yvf}.
The full-ensemble (subdivision) method~\cite{Welke:1989dr,LocalEnsemble,Zhang:1998tj,Molnar:2000jh}
 can reduce the causality violation,
but it alters event-by-event corrections and fluctuations and is
computationally expensive.

A Poincar\'e covariant Hamiltonian formalism for an $N$-interacting particle system
can be constructed based on the framework of constrained Hamiltonian
dynamics~\cite{Sudarshan:1981pp,Samuel:1982jk,Samuel:1982jn}
by extending the phase space from the non-relativistic $6N$ to 
$8N$ dimensions
to avoid the no-interaction theorem~\cite{Currie:1963dg,Currie:1963rw,Leutwyler:1965am},
where four-positions and four-momenta of particles are the dynamical variables.
The framework of constrained Hamiltonian dynamics%
~\cite{Sudarshan:1981pp,Samuel:1982jk,Samuel:1982jn}
was utilized to develop
a relativistic version of quantum molecular dynamics
(RQMD)~\cite{Sorge:1989dy,Maruyama:1991bp,Maruyama:1996rn,Isse:2005nk,Mancusi:2009zz,Marty:2012vs}
to describe the dynamics of interacting particles by mean-field potentials.
New versions of RQMD approaches are developed later
by implementing Lorentz-vector potentials~\cite{Nara:2019qfd,Nara:2020ztb,Nara:2021fuu}.


In this paper, we propose a Poincar\'e covariant cascade method
based on the framework of constrained Hamiltonian
dynamics~\cite{Sudarshan:1981pp,Samuel:1982jk,Samuel:1982jn}.
This method allows efficient numerical simulations of the collision processes
in $8N$-dimensional phase space, which is as efficient as the standard cascade method in a $6N$-dimensional
phase space.
In the constrained Hamiltonian dynamics,
the $N$-particle system is described in terms of $8N$-dimensional phase space.
To ensure that the dynamical system has the physical degrees of freedom of $6N$,
we impose $2N$ constraints.
We employ the same constraints as in Refs.%
~\cite{Maruyama:1996rn,Mancusi:2009zz,Marty:2012vs,Nara:2019qfd,Nara:2020ztb,Nara:2021fuu},
which were shown to describe interacting $N$ relativistic particles effectively
through potentials.
We compare our approach with other models
that employ different constraints~\cite{Sorge:1989dy,Maruyama:1991bp}
to investigate the difference and advantages of our proposed method.
Our covariant cascade method has been implemented
in the event generator JAM~\cite{Nara:1999dz,Nara:2019crj}
\footnote{The latest version of the code JAM2 is publicly available at
\url{https://gitlab.com/transportmodel/jam2}.}
to simulate high-energy nuclear collisions for a wide range of beam energy.


This paper is organized as follows: In Sec.~\ref{sec:6ncascade},
we summarize cascade methods in the $6N$-dimensional phase space.
In Sec.~\ref{sec:8ncascade},
a Poincar\'e covariant cascade method in $8N$-dimensional phase space
will be presented.
Section~\ref{sec:result} is dedicated to the comparison and validation of
different collision schemes.
We examine the collision rate and thermal spectra in a box calculation.
To investigate frame dependence,
we first analyze a one-dimensionally expanding system considering only elastic scatterings. Then, different collision schemes are applied to
collisions of two nuclei
in different computational frames,
including inelastic processes.
We also evaluate various collision schemes in
the $6N$-dimensional phase space~\cite{Zhang:1997ej,Bass:1998ca,Zhao:2020yvf}.
Our analysis reveals that some of the collision schemes provide a reasonable description of the
collision term when applied in the center-of-mass system under specific conditions.
In Sec.~\ref{sec:potential},
we will introduce new covariant equations of motion for a $N$-particle system
interacting through potentials, which is numerically efficient.
The conclusion is given in Sec.~\ref{sec:summary}.
Throughout this paper, we use the natural unit system of $c = k_B = 1$
and the sign convention of the metric $g_{\mu\nu}=\diag(+,-,-,-)$.

\section{Cascade methods in $6N$-dimensional phase space}
\label{sec:6ncascade}

We first summarize several cascade algorithms in the $6N$-dimensional phase space.
In these cascade models, the geometrical interpretation of the cross section
is commonly employed, where two particles will collide if the impact parameter
$b$ becomes smaller than the interaction range
specified by the cross section $\sigma$:
$b\leq\sqrt{\sigma/\pi}$.
The impact parameter is defined as
the minimum distance in the two-body center-of-mass system (c.m.s.).
Let us consider the collision of two particles with coordinates,
$q_1=(t_1,\bm{x}_1)$ and $q_2=(t_2,\bm{x}_2)$,
and momenta,
$p_1=(E_1,\bm{p}_1)$ and $p_2=(E_2,\bm{p}_2)$.
The Lorentz invariant expression of the squared impact parameter
is given by~\cite{Kodama:1983yk,Sasaki:2000pu,Hirano:2012yy}
\begin{equation}
  b^2 = \bm{x}_{\cm}^2 - \frac{(\bm{x}_{\cm}\cdot\bm{v}_{\cm})^2}{\bm{v}_{\cm}^2}
   = -q_T^2 + \frac{(q\cdot p_T)^2}{p_T^2},
\label{eq:impactpara}
\end{equation}
where $\bm{x}_{\cm}=\bm{x}_{\cm,1}-\bm{x}_{\cm,2}$ is the relative position
between the two particles,
and $\bm{v}_{\cm}=\bm{v}_{\cm,1}-\bm{v}_{\cm,2}$
is the relative velocity
in the two-body c.m.s., respectively.
The four-vectors $q_T$ and $p_T$ are the transverse relative position and momentum,
\begin{align}
  q &= q_1 - q_2, & p &= p_1 - p_2,\\
  q_T &= q - (q\cdot u)u, &
  p_T &= p - (p\cdot u)u,
\end{align}
where $u$ is 
the unit vector proportional to the total momentum of the two-body system
\begin{equation}
  u =\frac{p_1+p_2}{\sqrt{(p_1+p_2)^2}}.
\end{equation}
The time of the closest approach, $t_{\cm}$,
in the two-body c.m.s.,
\begin{align}
t_{\cm} &= t_{\cm,1} + \lambda(p_1,p_2,+q)(p_1\cdot u),\\
t_{\cm} &= t_{\cm,2} + \lambda(p_2,p_1,-q)(p_2\cdot u),
\end{align}
is Lorentz invariant quantities,
where
$t_{\cm,1}$ and $t_{\cm,2}$ denotes
the last collision times of the respective particles in the two-body c.m.s., and
\begin{equation}
 \lambda(p_1,p_2,q)= \frac{p_2^2(q\cdot p_1) - (p_1\cdot p_2)(q\cdot p_2)}
                         {(p_1p_2)^2 - p_1^2p_2^2}.
\end{equation}
The times of the closest approach $t_{c1}$ and $t_{c2}$ for the two colliding particles
in the computational frame
are obtained
by Lorentz transforming $t_{cm}$
(see Appendix~\ref{appendix:CD}):
\begin{align}
\begin{split}
 t_{c1} &= t_{1} + \lambda(p_1,p_2,+q) E_1,\\
 t_{c2} &= t_{2} + \lambda(p_2,p_1,-q) E_2.
\end{split}
\label{eq:colllab}
\end{align}
Although the times of the closest approach are simultaneous in the two-body c.m.s.,
they are, in general, different in a computational frame, which can lead to the violation of causality.
The two particles may collide when their impact parameter is less than the
interaction range $\sqrt{\sigma/\pi}$ and $t_{c1}>t_1$ and $t_{c2}>t_2$
to avoid backward collision in time.
We make a collision list by ordering all possible collision pairs.
The collision list is updated after every collision.
Thus, we need to define the collision ordering time~\cite{Zhang:1996gb,Zhao:2020yvf}.
The \textit{ordering time}, $t_o$, is used to order the collisions in the collision
list to determine which particle pair collides first
in the time-step--free cascade calculation.
The \textit{collision times}, $t_{\mathrm{coll1}}$ and $t_{\mathrm{coll2}}$, are
  the times
  at which the two particles collide, i.e., change their momentum
  (and species in inelastic scattering),
which may be different from the time of closest approach.

There is no unambiguous guideline to specify the ordering time.
The ZPC parton cascade model assumes the ordering time to be
the average of the times of the closest approach: $t_o=(t_{c1}+t_{c2})/2$,
and assumes that the collision times are
the same as the ordering time $t_\mathrm{coll1}=t_\mathrm{coll2}=t_o$.
This approach is referred to as collision scheme G (CS-G) in Ref.~\cite{Zhao:2020yvf}.
Another possible choice is to define the ordering time to be the minimum of
the times of the closest approach:
$t_o = \min(t_{c1},t_{c2})$, which predicts a less collision rate~\cite{Xu:2004mz}
(collision scheme A (CS-A)) with the collision times
being the same as the times of the closest approach:
$t_\mathrm{coll1}=t_{c1}$ and $t_\mathrm{coll2}=t_{c2}$.
In Ref.~\cite{Zhao:2020yvf}, a collision scheme, in which
the collision times and the collision ordering
time are chosen by the minimum: $t_\mathrm{coll1}=t_\mathrm{coll2}=t_o = \min(t_{c1},t_{c2})$
(collision scheme B (CS-B)),
accurately describes the collision rate and the equilibrium momentum distribution
in a box calculation.
The distinct difference in the collision schemes is that
two particles collide simultaneously in their two-body c.m.s. in CS-A,
while in CS-B and CS-G, the collision time of two colliding particles is the same in the computational
frame.

There are problems in these schemes due to the different times for the
closest approach in the computational frame.
For example, in collision scheme CS-A, the particle with a larger collision time
would not collide in the duration between the collision times
$|t_{c1}-t_{c2}|$ to avoid non-causal collisions,
which reduces the collision rate~\cite{Zhang:1998tj}.
We note that particles do not collide at the distance of the closest approach
in collision schemes CS-B and CS-G\@.
In these collision schemes, to avoid non-causal collisions,
we impose the additional conditions:
$t_\mathrm{coll1}>t_1$ and $t_\mathrm{coll2} > t_2$.

As an alternative approach,
the proper time interval $\delta\tau_i(j)$
from the last collision time $\tau_i$ to the next collision time $\tau_{ci}(j)$
for the collision of particles $i$ with particle $j$ at the closest approach
\begin{align}
\delta\tau_i(j) &=\tau_{ci}(j) - \tau_i = \lambda(p_i,p_j,q_i-q_j)\sqrt{p_i^2},
\label{eq:kodama}
\end{align}
are considered for the ordering of two-body collisions in Ref.~\cite{Kodama:1983yk}.
It is argued that
in this approach, if only causal collisions are required,
the number of collisions is largely reduced and may lead to the underestimation
of the collision rate~\cite{Kortemeyer:1995di}.

To avoid the problems of collision ordering,
one can introduce the time step, and collision pairs are randomly selected
by using the predicted collision times in Eq.~\eqref{eq:colllab}
~\cite{Wolf:1990ur}.
We note that the Lorentz transformation of
Eq.~\eqref{eq:kodama} to the computational frame  yields Eq.~\eqref{eq:colllab}.
In this approach, a sufficiently small time-step size should be selected
to avoid the artifact of the order of the collision and decay
sequence~\cite{TMEP:2019yci}.
In addition, the comparisons of $N(N-1)/2$ pairs in a $N$-particle system have to be made at each
time step, which leads to the increasing computational time proportional to $N^2$.

In UrQMD~\cite{Bass:1998ca} and SMASH~\cite{SMASH:2016zqf},
the collision ordering time and the collision time
are defined as the time of the closest approach in the computational reference frame,
\begin{equation}
 t_\mathrm{coll}^\mathrm{(ref)} = t_1 -\frac{[\bm{x}_1(t_1) - \bm{x}_2(t_1)]\cdot(\bm{v}_1-\bm{v}_2)}
             {(\bm{v}_1-\bm{v}_2)^2},
\end{equation}
where coordinates and velocities $\bm{v}_1$ and $\bm{v}_2$ are taken in the
computational frame.
It should be noted that the impact parameter is still defined in the two-body c.m.s. frame.
Let us call the collision scheme with this collision ordering CS-O.
This collision scheme does not have problems arising from the difference
in collision times.

As one possible way to recover the Lorentz covariance,
the full-ensemble method has been proposed%
~\cite{Welke:1989dr,LocalEnsemble,Zhang:1998tj,Molnar:2000jh},
in which particles are over-sampled by
the factor of $N_\mathrm{test}$: $n\to nN_\mathrm{test}$ and the cross section is
reduced by $\sigma\to \sigma/N_\mathrm{test}$.
As the density $\rho$ scales as $N_\mathrm{test}$,
the mean free path remains the same $\ell \sim (\rho\sigma)^{-1}$.
Since the collision distance scales as $d=\sqrt{\sigma/\pi}\propto 1/\sqrt{N_\mathrm{test}}$,
the collision occurs at the same space-time point in the limit of $N_\mathrm{test}\to\infty$
so that the collision term of the Boltzmann equation is recovered.
However, this method destroys the event-by-event fluctuations and correlations.
Furthermore, it requires intensive computational time.

As an alternative method,
the ``local-ensemble method'' \cite{Danielewicz:1991dh,LocalEnsemble,Cassing:2001ds},
which is also called the ``stochastic method''~\cite{Xu:2004mz}
was developed based on the covariant collision rate.
We first divide the space into a set of small volumes and introduce over-sampled test particles by the factor $N_\mathrm{test}$.
When the number of collisions occurring in volume $dV$ in a time interval $dt$ is
given by~\cite{Landau:1975pou}
\begin{align}
 d\nu &= \sigma v_M n_1n_2dVdt, &
    v_M &= \frac{\sqrt{(p_1\cdot p_2)^2 - p_1^2 p_2^2}}{p_1^0 p_2^0},
\end{align}
one obtains the collision probability in volume $dN$ and time interval $dt$ as
\begin{equation}
P_{22} =  \frac{d\nu}{dN_1dN_2} = \frac{\sigma}{N_\mathrm{test}} v_M \frac{dt}{dV},
\end{equation}
where $n_i = dN_i/dV$.
In each time step, 
the collision pairs are randomly chosen from the volume.
In the limits of $dV\to0$, $dt\to0$, and $N_\mathrm{test}\to\infty$,
the collision rate converges to the exact solutions of the Boltzmann
equation.
This approach has the advantage that
multi-particle collisions such as $m\to n$
processes can be simulated in a straightforward way%
~\cite{Cassing:2001ds,Staudenmaier:2021lrg,Coci:2023daq}.
The stochastic method generally requires a large number of test particles
to ensure the accurate solution of the Boltzmann equation.
The cell length should be smaller than the mean free path to reduce numerical
artifacts.
This method was first used to solve the BUU transport model, which solves
the time evolution of the one-particle phase space function.

\section{Poincar\'e Covariant cascade models}
\label{sec:8ncascade}

We now consider the Poincar\'e covariant cascade method in the $8N$-dimensional phase space
in terms of four-vectors of the particle positions and momenta
to construct
a covariant cascade method.

\subsection{Covariant cascade method from the constrained Hamiltonian formalism}

According to Dirac's constrained Hamiltonian formalism,
the Hamiltonian is given by a linear combination of $2N-1$ constraints,
\begin{equation}
H=\sum_{i=1}^{2N-1} u_i \phi_i.
\end{equation}
The trajectories of four-positions $q_i(\tau)$ and four-momenta $p_i(\tau)$ are
parametrized by the Poincar\'e invariant evolution parameter $\tau$.
The equations of motion are given by
\begin{eqnarray}
\frac{dq^\mu_i}{d\tau} &=& [H,q_i^\mu]
 = \sum_{j=1}^{2N-1} u_j\frac{\partial\phi_j}{\partial p_{i\mu}},
\label{eq:EOM1}\\
\frac{dp^\mu_i}{d\tau} &=& [H,p_i^\mu]=
 -\sum_{j=1}^{2N-1} u_j\frac{\partial\phi_j}{\partial q_{i\mu}},
\label{eq:EOM2}
\end{eqnarray}
where the Poisson brackets are defined as
\begin{equation}
[A,B] = \sum_k \left(
 \frac{\partial A}{\partial p_k}\cdot
 \frac{\partial B}{\partial q_k} -
 \frac{\partial A}{\partial q_k}\cdot
 \frac{\partial B}{\partial p_k}
   \right).
\end{equation}

In the cascade method, particles travel on the straight-line
trajectory as if they are free most of the time,
and the collision takes place at discrete points in $\tau$.
Thus, we take
the following mass-shell constraints:
\begin{equation}
H_i = \phi_i = p_i^2 - m_i^2, \quad i=1,\ldots,N\,.
\label{eq:massshell}
\end{equation}
The remaining $N$ constraints, which we call \textit{time fixation}, fix the times of particles.
Different time fixations will be discussed in the following subsections.

The Lagrange multipliers $u_i$ are determined by requiring the invariance of the
constraints during the time evolution:
\begin{equation}
\frac{d\phi_i}{d\tau} = [H,\phi_i] + \frac{\partial\phi_i}{\partial\tau}
= \sum_{j=1}^{2N-1}C_{ij}u_j
 + \frac{\partial\phi_{i}}{\partial\tau} = 0,
\end{equation}
where $C_{ij}=[\phi_j,\phi_i]$.
Since the mass shell constraints commute $[H_i,H_j]=0$,
the Lagrange multipliers can be obtained by
solving $N\times N$ system of linear equations for $u_j$.

In the following, we will discuss different realizations of cascade models
within the constrained Hamiltonian formulation by employing different time
fixations.

\subsection{Sorge's constraints}

One of the main requirements in relativistic particle dynamics
is the cluster separability.
The cluster separability means that
when the Minkowski distance between two clusters of particles is space-like and sufficiently large,
they do not interact with each other.

Based on the idea~\cite{Samuel:1982jn},
Sorge proposed the following time constraints~\cite{Sorge:1989dy},
which  satisfy the cluster separability,
\begin{equation}
 \chi_i = \sum_{i\neq j} G_{ij} u_{ij}\cdot q_{ij}
     =0, \quad i=1,\ldots,N-1, \label{eq:SorgeConst}
\end{equation}
where
$p_{ij}=p_i+p_j$,
$q_{ij}=q_i - q_j$, and
$u_{ij} =  p_{ij}/\sqrt{p_{ij}^2}$.
The evolution parameter $\tau$ may be fixed by
the gauging condition~\cite{Maruyama:1991bp} that
equates the average time of particles in the global c.m.s.
to $\tau$:
\begin{equation}
 \chi_N = Q\cdot U - \tau=0,
\end{equation}
where
\begin{align}
Q &= \frac{1}{N}\sum_{i=1}^N q_i, &
P &=  \sum_i p_i, &
U &= \frac{P}{\sqrt{P^2}}.
\end{align}
The weight function $G_{ij}$
only depends on the Minkowski distance $q_{ij}$ and
is chosen as
\begin{equation}
 G_{ij}= \frac{\exp \left(q_{ij}^2/L\right)}{q_{ij}^2/L},
\label{eq:Gij}
\end{equation}
where we use $L=8$ fm$^2$.
In these constraints, the times of the interacting two particles in their c.m.s. 
are equal in the dilute limit.
The terms $1/q_{ij}^2$ are introduced
to keep the space-like distance between two particles to
maintain the causality.

Numerical implementation of the constraints~\eqref{eq:SorgeConst}
is quite challenging for the simulation of realistic collisions.
At every two-body collision and decay, time constraints~\eqref{eq:SorgeConst}
break. Thus, one needs to solve the non-linear equations in $N$ unknowns to recover
the time constraints after every collision or decay.
It is too time-consuming to perform such numerical simulations on a current computer.
In addition, the standard Newton-Raphson method often does not find the solution.
Therefore, we simulate an approximate solution of the constraints by
introducing time steps, and we try to recover
the time constraints at every time step.
We recognize clusters at every time step and
solve these constraints for each cluster.
Furthermore, we approximate the solution in only one iteration in the
Newton-Raphson method. 
We also assume that the Lagrange multiplier $u_i$ remains
the same during each time step after two-body collisions.
It should be mentioned that in the actual numerical simulations,
the constraints given by Eq.~\eqref{eq:SorgeConst} occasionally result
in a negative Lagrange multiplier $u_i$, 
which means that the $i$th particle  moves backward in time.
In such cases, we assume that the Lagrange multiplier is equal to
zero, $u_i=0$.


\subsection{Poincar\'e covariant parton cascade model}

The Poincar\'e covariant parton cascade (PCPC)
~\cite{Borchers:2000wf}
is an alternative approach
for a manifestly covariant formulation
of the cascade model. This model is based on a Hamiltonian formulation
that allows $N$ particles to move freely in an $8N$-dimensional phase space.
The Hamiltonian for the free particle system~\cite{Peter:1994yq,Behrens:1994yr}
is given by\footnote{This is actually a constraint and different from the usual Hamiltonian in the sense
of total energy.}
\begin{equation}
 H = \sum_{i=1}^N \frac{p_i^2 - m_i^2}{2m_i},
\end{equation}
and Hamilton's equations of motion
specify the world lines of particles
\begin{equation}
 \frac{dq_i}{d\tau} = \frac{p_i}{m_i} \equiv u_i,
\label{eq:pcpc}
\end{equation}
where the four-position vector $q_i$ is a function of
the Poincar\'e invariant parameter $\tau$.
In the case of a massless particle system,
the Hamiltonian~\cite{Kurkela:2022qhn}
\begin{equation}
 H = \sum_{i=1}^N \frac{\lambda}{2} p_i^2
\end{equation}
gives the equations of motion
\begin{equation}
 \frac{dq_i}{d\tau} = \lambda p_i,
\label{eq:pcpc0}
\end{equation}
where $\lambda$ is an arbitrary constant.
The collision can happen when the two-body distance in the center-of-mass
system becomes the closest.

Let us consider this approach based on the constrained Hamiltonian dynamics.
The first $N$ constraints are the same as the free mass-shell constraints.
To find the time constraints,
by multiplying $u_i$ by both sides of Eq.~\eqref{eq:pcpc},
one obtains
\begin{equation}
 u_i\cdot \frac{dq_i}{d\tau}= 1.
\label{eq:pcpcq}
\end{equation}
After integrating Eq.~\eqref{eq:pcpcq} over $\tau$,
the time fixation constraints are expected to be
\begin{eqnarray}
\chi_{i} &=& u_i\cdot [q_i(\tau)-q_i(\tau_{0i})] - \tau + \tau_{0i},
\label{eq:constpcpc}
\end{eqnarray}
where $i$ runs over $i=1,\ldots,N$.
Indeed, the Lagrange multipliers are given by
$u_i=1/(2p_i\cdot u_i)=1/2m_i$ and we obtain the equations of motion~\eqref{eq:pcpc}.
One can choose the initial coordinate $q_i(\tau_{0i})$  and evolution parameter
$\tau_{0i}$ for the particle arbitrarily at the collision point, thus there are no constraints in
Eq.~\eqref{eq:constpcpc}.
Moreover, the time constraints do not depend on other particles.

Similarly, for the massless particle case,
Eq.~\eqref{eq:pcpc0} implies
\begin{equation}
p_i\cdot\frac{dq_i}{d\tau}=0, \quad i=1,\ldots,N.
\end{equation}
Thus, the equations of motion for massless particles
may be obtained from the constraints
\begin{align}
\phi_i &= p_i^2, \quad i=1,\ldots,N,\\
\phi_{N+i} &= p_i\cdot [q_i(\tau)-q_i(\tau_{0i})],
  \quad i=1,\ldots,N-1, \\
\phi_{2N} &= Q\cdot \frac{P}{P^2}  - \lambda \tau,
\end{align}
where $Q=\sum_{i=1}^N q_i$ and $P=\sum_{i=1}^N p_i$.
The last constraint implies $t_\mathrm{cm}=\lambda E_\mathrm{cm}\tau$,
where $t_\mathrm{cm}=Q^0$ and $E_\mathrm{cm}=P^0$ in the global c.m.s.
One finds that the Lagrange multipliers are the same for all particles: $u_i=\lambda/2$,
and the equations of motion are identical to Eq.~\eqref{eq:pcpc0}.

In the PCPC approach,
the times of the particles elapse in proportion to their energies, and
there is no time relation between particles.
Times of the particles are randomly spread at later times.
Consequently,
particle distances can be time-like, which violates the causality
because a particle can collide with an absolute-future particle.
In a box simulation,
ignoring such acausal collisions
leads to a significant reduction in collision rate at late times.
However, when simulating the collision of two nuclei,
these effects may not be as relevant at a certain level
because the momenta, and thus the times of particles are similar when they are close to each other
in the coordinate space.

\subsection{A new covariant cascade method}

In this section, we introduce our covariant cascade method
based on the constrained Hamiltonian dynamics
taking the constraints used in Refs.~\cite{Sudarshan:1981pp,Maruyama:1996rn}
for the potential interactions.
Specifically, we apply
the mass shell constraints~\eqref{eq:massshell} and
the following time constraints 
to formulate a cascade method for the first time:
\begin{eqnarray}
\phi_{N+i} &=& \chi_i= \hat{a}\cdot (q_i - q_N), \quad i=1,\ldots,N-1,
\label{eq:timefix}\\
\phi_{2N} &=& \chi_N = \hat{a}\cdot q_N - \tau,
\label{eq:timefix2}
\end{eqnarray}
where $\hat{a}$ is a Lorentz vector, which acts as a timekeeper.
In Refs.~\cite{Samuel:1982jk,Marty:2012vs},
$\hat{a}$ is defined as
$\hat{a}^\mu=P^\mu/(P_\mu P^\mu)^{1/2}$, where $P^\mu$ is the total momentum of the system. This choice equates the times of all particles
in the global c.m.s\@. In other words, the evolution parameter $\tau$ is interpreted
as the time as measured in the global c.m.s\@.
Another way to specify $\hat{a}$ is to introduce a freely moving dummy particle
and to specify $\hat{a}$ using the momentum of the dummy particle~\cite{Sudarshan:1981pp}.
If we choose the rest frame of the dummy particle as the global c.m.s.,
the two models become identical.
We note that a different choice of timekeeper yields a different physical
system of interacting particles.
Throughout this work, we define $\hat{a}=(1,0,0,0)$ in the global c.m.s\@.
We should note that the four-vector $\hat{a}$ is subject to the Poincar\'e transformation
in switching to another computational frame
so that the Poincar\'e covariance of the model is maintained.

The Lagrange multipliers are solved to be
$u_i = 1/(2\hat{a}\cdot p_i)$,
and the equations of motion are obtained as
\begin{equation}
\frac{dq_i}{d\tau} = \frac{p_i}{\hat{a}\cdot p_i}.
\end{equation}
Note that $\hat{a}\cdot p_i$ is the energy of the $i$th particle
in the frame in which $\hat{a}=(1,0,0,0)$.  The time coordinate of the $i$th
particle is identical to the evolution parameter $q_i^0=\tau$
in this frame, but it is different in other frames.
We note that the physical velocity of the particle
\begin{equation}
\frac{d\bm{q}_i}{dq_i^0}
= \frac{d\bm{q}_i}{d\tau}\frac{d\tau}{dq_i^0}
= \frac{\bm{p}_i}{p_i^0}
\end{equation}
is kept to be less than the speed of light.

Our approach satisfies the cluster separability
in the sense that
independent clusters do not interact with each other
since
the timekeeper $\hat{a}$ is a constant vector
after specifying the computational frame.
However,
when a cluster is separated into clusters, the c.m.s. of each
cluster is different from the original frame.
This implies that to be consistent with the simulations started from the separated state as an initial condition,
the different $\hat{a}$ must be reassigned to each cluster,
which is not included in our constraints.
We will discuss this issue with a different model in Sec.~\ref{sec:cluster}.

\subsection{Closest distance approach in $8N$-dimensional phase space}

Let us now consider a covariant cascade procedure.
The main interest is to determine the collision point of two particles in a
covariant way.  This is straightforward in the $8N$-dimensional phase-space approach.
In the cascade model, particles move in straight lines and change their momentum only by collisions or decays.
The distance squared between two particles is defined as the distance
in the two-body c.m.s.:
\begin{equation}
  d^2 = -q^2 + \frac{(q\cdot P)^2}{P^2},
\end{equation}
where $q=q_1 - q_2$ and $P=p_1 + p_2$.
The time of the closest distance may be obtained by the condition
\begin{equation}
  -\frac{1}{2}\frac{\partial d^2}{\partial \tau}
 = q_T\cdot v
 =  q \cdot v_T
 = 0,
\end{equation}
where
$v = \frac{dq}{d\tau}$ is the relative velocity, and
\begin{equation}
 v_T = v - \frac{(v\cdot P)}{P^2}P
\end{equation}
is the transverse relative velocity of the colliding two particles.
Applying the closest distance condition $q\cdot v_T=0$ to the equations of motion of two particles,
\begin{equation}
q_i(\tau)  = q_i(\tau_i) + v_i\,(\tau - \tau_i), \quad i=1,2,
 \label{eq:ceom}
\end{equation}
where $v_i=p_i/(\hat{a}\cdot p_i)$ is a constant velocity in our model,
we get the time of the closest approach $\tau_\mathrm{coll}$ as
\begin{align}
\tau_\mathrm{coll} - \tau_i &= -\frac{ (q_1(\tau_i)-q_2(\tau_i))\cdot v_T}{v_T^2}, \quad i=1,2.\label{eq:ctime1}
\end{align}
The two-body collision will occur at $\tau_\mathrm{coll}$, and it is ordered
in $\tau$ in a frame-independent way.
In this way, we avoid the frame dependence of the ordering time.


The closest distance $b^2$, i.e., squared impact parameter, may be obtained by
\begin{equation}
  b^2 = -q_\mathrm{c}^2 + \frac{(q_\mathrm{c}\cdot P)^2}{P^2},
\label{eq:impact}
\end{equation}
where $q_\mathrm{c} = q_1(\tau_\mathrm{coll})-q_2(\tau_\mathrm{coll})$ is the relative distance when the collision takes place.
We note that $b^2$ is identical to the previously presented impact parameter%
~\eqref{eq:impactpara} (see Appendix~\ref{appendix:IP}).

\section{Results}
\label{sec:result}

In the following, we validate the thermal distribution and collision rate
of our covariant scheme
within a box simulation
assuming massless particles and elastic scattering.
We compare the simulation results with analytical expressions.
Next, we examine the frame dependence of the results for a one-dimensionally expanding system.
Furthermore, we present simulation results of nucleus-nucleus collisions
from compression stages to expansion stages, including inelastic processes.

In Table~\ref{tab:CS},
we summarize the collision schemes that will be compared in this section.

\begin{table}[tbhp]
\caption{Different collision schemes in the closest approach.
In the covariant collision scheme, collisions happen
at a single collision time specified by the Lorentz invariant evolution parameter
$\tau_\mathrm{coll}$.
In CS-A, CS-B, and CS-G, the times of the closest approach for the two colliding particles
are computed in the two-body center-of-mass system (c.m.s.), then these times
are transformed to the computational frame, which is represented by $t_{c1}$ and $t_{c2}$.
In CS-O,
the collision time $t_\mathrm{coll}^\mathrm{(ref)}$ is
the time of the closest approach computed in the computational frame.
The frame where the impact parameter $b$ is defined
is listed.
Other collision schemes are examined in Ref.~\cite{Zhao:2020yvf}.
We list CS-M in Ref.~\cite{Zhao:2020yvf} to avoid confusion with CS-O.
}
\centering
\begin{tabular}{l|ccc}
\hline
\hline
& collision time ~~~& ordering time~~~ & frame of $b$ \\
\hline
\hline
Covariant & $\tau_\mathrm{coll}$     &  $\tau_\mathrm{coll}$  & two-body c.m.s\\
CS-A & $t_{c1}$ and $t_{c2}$     &  $\min(t_{c1},t_{c2})$ & two-body c.m.s.\\
CS-B & $\min(t_{c1}$, $t_{c2})$  &  $\min(t_{c1}, t_{c2})$ & two-body c.m.s.\\
CS-G & $(t_{c1}$ + $t_{c2})/2$  &  $(t_{c1}+t_{c2})/2$  & two-body c.m.s.\\
CS-O & $t_\mathrm{coll}^\mathrm{(ref)}$  &  $t_\mathrm{coll}^\mathrm{(ref)} $
& two-body c.m.s.\\
CS-M & $t_\mathrm{coll}^\mathrm{(ref)}$  &  $t_\mathrm{coll}^\mathrm{(ref)} $
& computational frame\\
\hline
\end{tabular}
\label{tab:CS}
\end{table}

\subsection{Box calculations}

We first examine energy distribution in a box calculation.
We assume massless gluons, elastic collisions with the cross section
of $\sigma=2.6$ fm, and isotropic scattering~\cite{Zhao:2020yvf}.
The coordinates of the particles are uniformly distributed,
and the momenta of the particles are initialized as
\begin{equation}
 \frac{dN}{Ndp_Tdp_z} = \delta(p_T - 1.5\,\mathrm{GeV})\delta(p_z)
\end{equation}
with the total number of particles $N=5400$ in the box size
of 6 fm$\times$6 fm$\times$6 fm.
At late times, the distribution should obey the Boltzmann form
\begin{equation}
 \frac{dN}{NE^2dE} = \frac{1}{2T^3}e^{-E/T}.
\label{eq:BoltzmannDist}
\end{equation}

\begin{figure}[tbh]
\includegraphics[width=8.5cm]{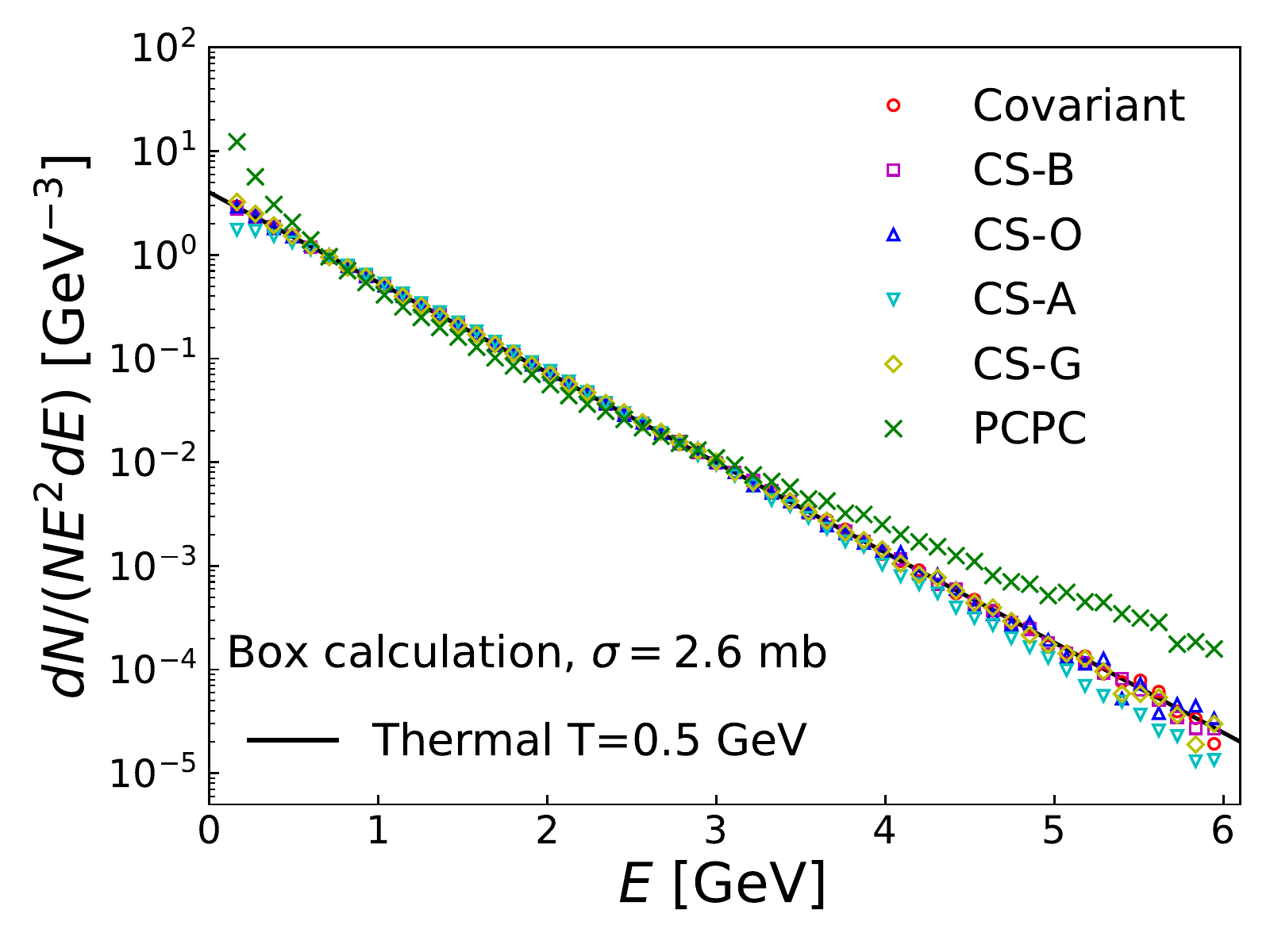}%
\caption{
Energy distributions are compared for different collision schemes.
The size of the box is 6 fm$\times$6 fm$\times$6 fm.
The cross section for the massless gluons is set to 2.6 mb,
and an isotropic angular distribution is assumed.
The line represents the thermal distribution of temperature $T=0.5$~GeV\@.
}
\label{fig:dnde_box}
\end{figure}

Figure~\ref{fig:dnde_box} compares the final energy distributions
among different collision schemes in a box calculation together with the
theoretical curve~\eqref{eq:BoltzmannDist} with the temperature $T=0.5$~GeV\@.
The final energy distributions are obtained after the global time of 5 fm/$c$.
The energy distribution from collision scheme CS-G, which uses the average collision time for the ordering time,
deviates from the thermal distribution.
The results from the other collision schemes are in good agreement
with the expected thermal distribution except for the PCPC approach.
As explained in the previous section, the PCPC evolution does not have
any time correlation between particles; particle times
spread without any restriction, which leads to timelike distances
between particles.
This might be the main reason for unphysical distribution from the PCPC approach.
We note that the results of
the $6N$-dimensional phase-space cascade schemes of CS-B and CS-G
are consistent with the results in Ref.~\cite{Zhao:2020yvf}.

\begin{figure}[tbh]
\includegraphics[width=8.5cm]{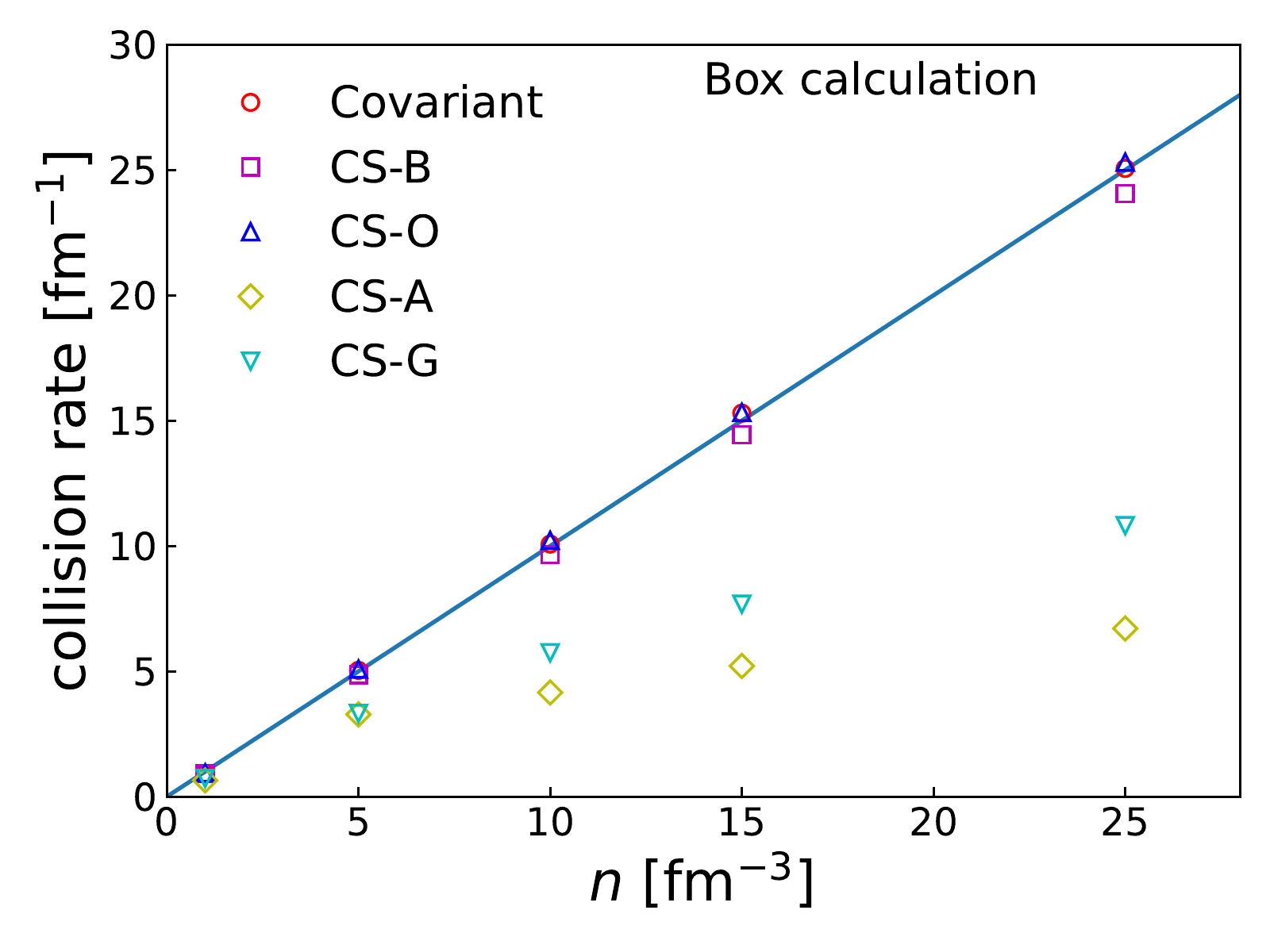}%
\caption{
Collision rate per collision pair as a function of density.
The size of the box is 6 fm$\times$6 fm$\times$6 fm.
The cross section for the massless gluons is set to 10 mb,
and an isotropic angular distribution is assumed.
The line corresponds to the analytical expectation $\sigma n$.
}
\label{fig:ncoll_box}
\end{figure}

In Fig.~\ref{fig:ncoll_box}, the collision rates per collision pair
in a box calculation are plotted
as functions of density $n$ for different collision schemes.
We use the cross section of $\sigma=10$ mb and isotropic scattering.
Collision schemes CS-A and CS-G predict much lower collision rates
than the analytical expectation $\sigma n$, which is consistent with the
result in Refs.~\cite{Xu:2004mz,Zhao:2020yvf};
the collision rate is significantly suppressed in schemes CS-A and CS-G
when the mean free path is smaller than the interaction range
because the number of non-causal collisions increases~\cite{Zhang:1998tj}.
The other collision schemes are consistent with the analytical expectation, including
our covariant cascade method.
As discussed in detail in Ref.~\cite{Xu:2004mz},
the difference in the collision times in the computational frame
in collision schemes CS-A and CS-G leads to
the lower collision rate.
Let us consider the collision of particles 1 and 2 with $t_1$ and $t_2$ being
the times when the last collision of particles 1 and 2 happened.
The times of the closest approach are denoted by $t_{c1}$ and $t_{c2}$.
In the scheme CS-A, the collision ordering time
is defined as the smaller one of the two collision times: $t_o = \min(t_{c1},t_{c2})$.
In the situation $t_1 < t_{c1} < t_2 < t_{c2}$,
the collision between particles 1 and 2 occurs at $t=t_{c1}$,
which is earlier than $t_2$, the time of the collision between particle 2 and another.
To forbid such non-causal collisions, we need an additional
condition $t_o > t_i,~i=1,2$, which reduces the number of collisions
since possible collisions of particle 2 during the time interval
$|t_{c1}-t_{c2}|$ are not considered.
In contrast, collision scheme CS-B assumes the collision time and the ordering
time to be the smallest one: $t_o=t_\mathrm{coll1}=t_\mathrm{coll2}=\min(t_{c1},t_{c2})$,
which
does not suppress possible collisions in the interval $|t_{c1}-t_{c2}|$.
The collision rate
in the collision scheme CS-B is close to the analytical one
in the box simulation as found in Ref.~\cite{Zhao:2020yvf}.
The collision scheme CS-O, in which the collision times of particles and the
collision ordering time are the same and are determined in the computational frame,
correctly reproduces the analytical value since there is no problem arising
from the difference in the collision times.

\subsection{One-dimensionally expanding systems}

To study the frame dependence of cascade results,
we examine a one-dimensionally expanding system that is motivated by
the initial conditions at RHIC (Relativistic Heavy Ion Collider).
For this purpose, we adopt the same initial condition as in Ref.~\cite{Zhang:1996gb}.
Specifically, we consider a system consisting of
$N=4000$ gluons that are uniformly distributed
within a spacetime rapidity interval $\eta\in[-5,5]$.
The transverse distribution of particles is assumed to be a disk of radius 5 fm.
The momentum distribution of the particles is taken from the Boltzmann distribution
with the temperature of 0.5~GeV and boosted with the velocity
$\beta=\tanh\eta_i$, where $\eta_i$ is the spacetime rapidity of the $i$th
particle.
The times and longitudinal positions of particles are then obtained,
assuming the formation time of $\tau_0=0.1$ fm/c,
\begin{align}
 t_i &= \tau_0\cosh\eta_i, &
 z_i &= \tau_0\sinh\eta_i.
\label{eq:iniBJ}
\end{align}
The transverse positions $\bm{x}_{Ti}$ are propagated up to the formation
time, $\bm{x}_{Ti}(t_i) = \bm{x}_{Ti}(0) + \bm{v}_{Ti}t_i$.
To study the frame dependence, we boost the above initial condition defined in
the global c.m.s.
(the \textit{collider frame})
using the velocity $\beta=\tanh\eta_0$ with $\eta_0=5$
(the \textit{target frame}).

\begin{figure}[tbh]
\includegraphics[width=8.5cm]{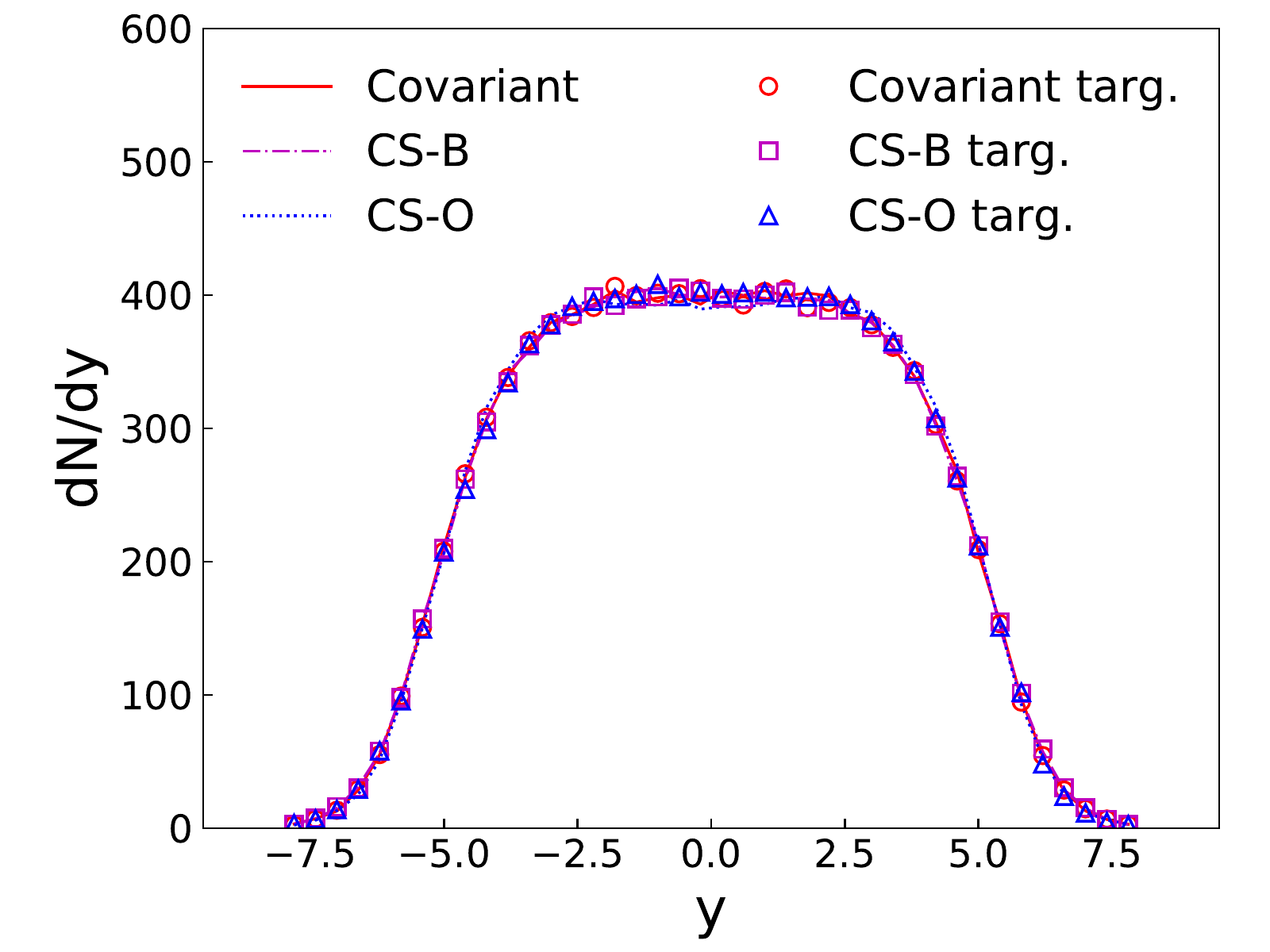}%
\caption{
Rapidity distributions are compared for different collision schemes.
The cross section for the massless gluons is set to 3.5 mb,
and an isotropic angular distribution is assumed.
The solid, dashed-dotted, and dotted lines correspond to the results
from the covariant collision scheme, CS-B, and CS-O in the collider frame.
The results from the target frame are shown by the circles, squares, and
triangles for the covariant collision scheme, CS-B, and CS-O, respectively.
}
\label{fig:dndy400}
\end{figure}

In Fig.~\ref{fig:dndy400}, we compare the rapidity distributions with
different collision schemes for different frames.
The solid, dashed-dotted, and dotted lines correspond to the results
from the covariant cascade, CS-B, and CS-O, respectively.
The corresponding results from the calculations in the target frame
are expressed by symbols.
We do not show collision schemes CS-A and CS-G because the frame dependencies
turned out to be the same as that of scheme CS-B\@.
In the one-dimensionally expanding system with elastic collisions,
we do not see frame dependence even for the non-covariant collision schemes.
In Ref.~\cite{Zhang:1996gb}, strong frame dependence is observed in the
collision scheme in which both the collision frame and the defining frame
of the impact parameter of the two-body scattering are the computational frame.
In collision scheme CS-O, the time of the closest approach is computed in the
computational frame while the impact parameter is defined in the two-body c.m.s. 
frame of colliding pair. This is the reason why the frame dependence is
greatly suppressed in collision scheme CS-O\@.


\begin{figure}[tbh]
\includegraphics[width=8.5cm]{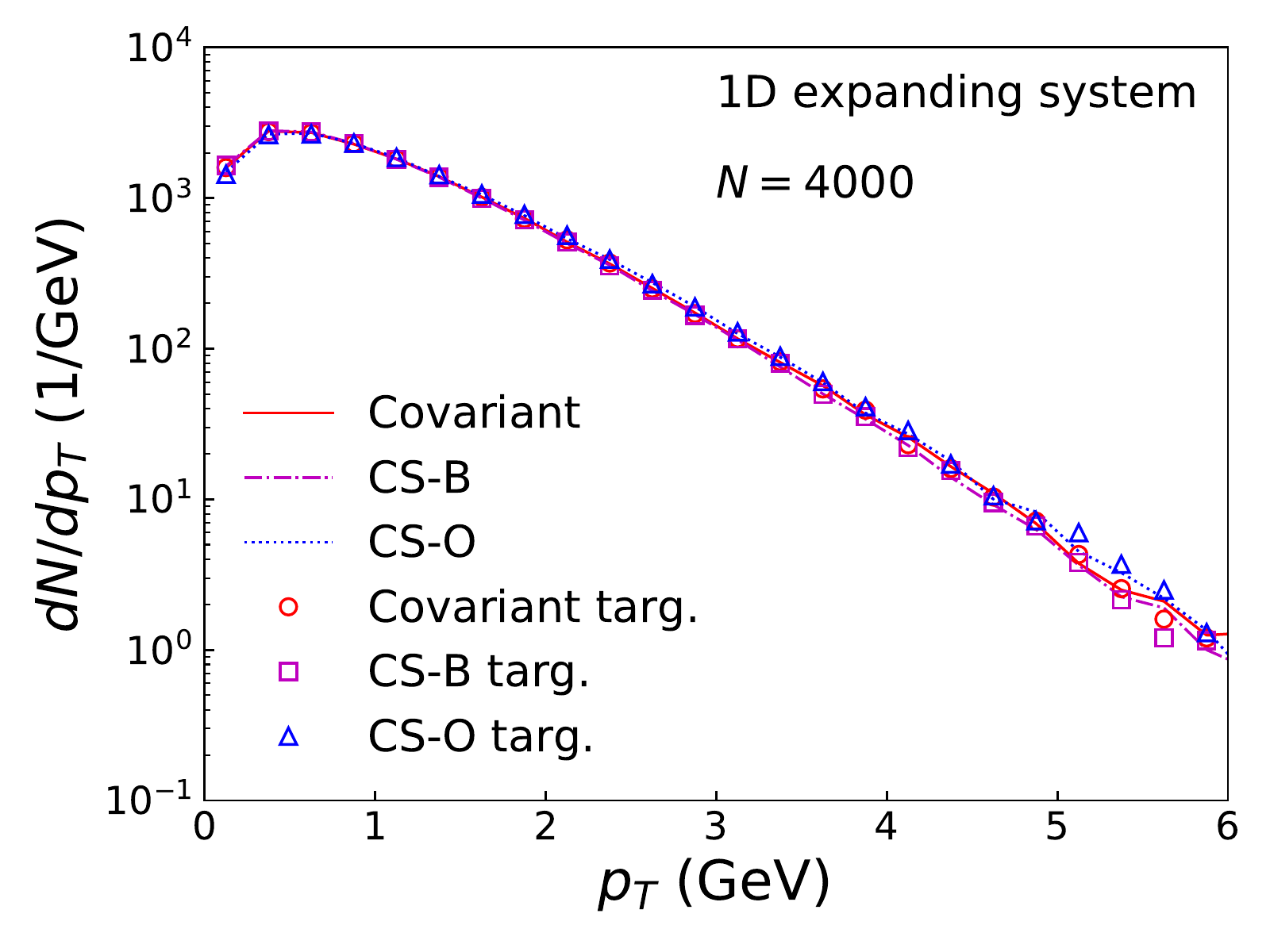}%
\caption{Transverse momentum distributions are compared for different collision schemes
The cross section for the massless gluons is set to 3.5 mb,
and an isotropic angular distribution is assumed.
The meaning of the lines and symbols is the same as in Fig.~\ref{fig:dndy400}.
}
\label{fig:dndpt400}
\end{figure}

The transverse momentum distributions are compared in Fig.~\ref{fig:dndpt400}.
The results from three collision schemes in the collider frame are indistinguishable.
We see a slight enhancement of the transverse energy in
collision scheme CS-O in the target frame.

\begin{figure}[tbh]
\includegraphics[width=8.5cm]{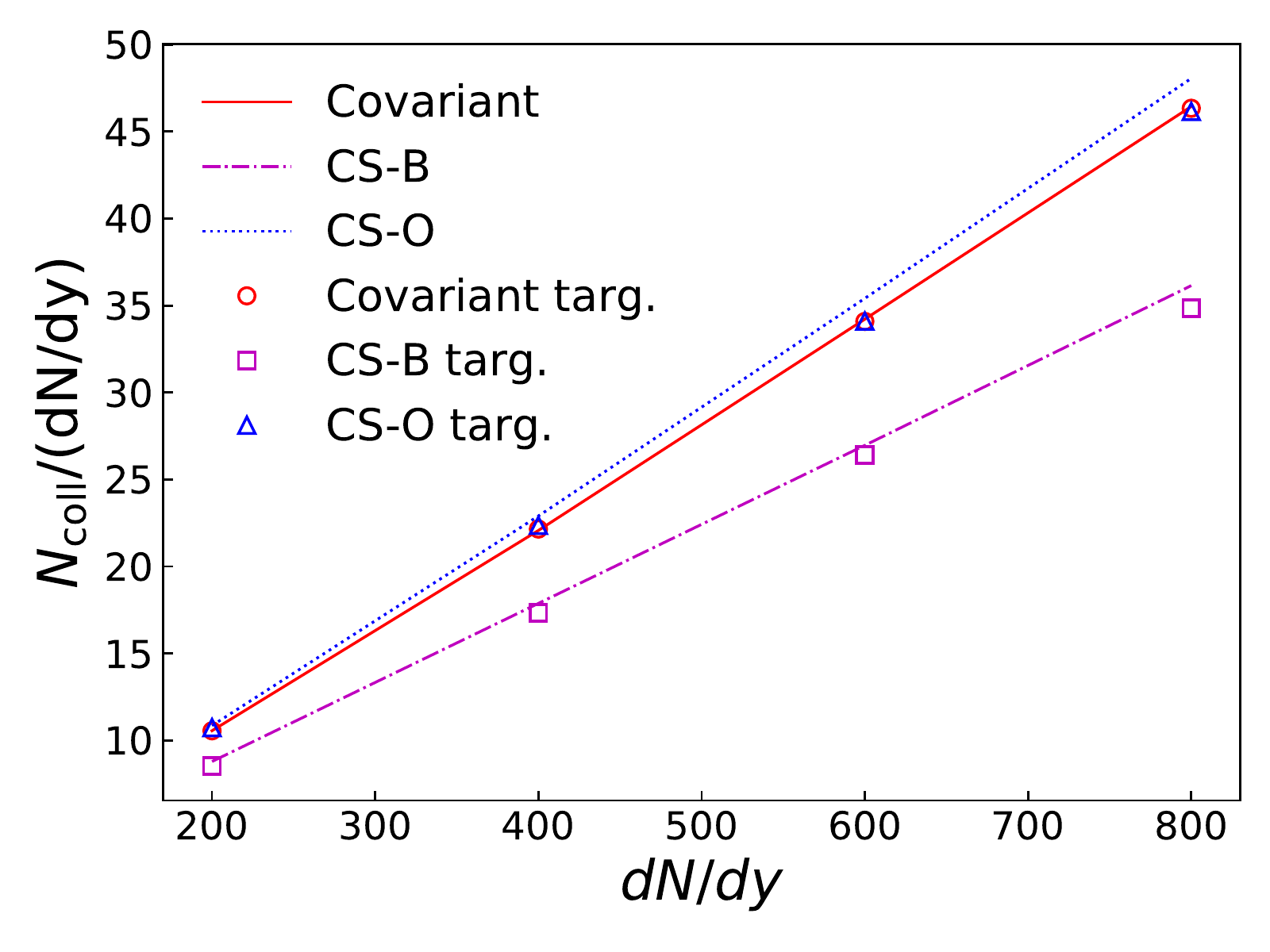}%
\caption{The total number of collisions normalized by the $dN/dy$ as a function of $dN/dy$
for the one-dimensionally expanding initial condition.
The cross section for the massless gluons is set to 3.5 mb,
and an isotropic angular distribution is assumed.
The meaning of the lines and symbols are the same as in Fig.~\ref{fig:dndy400}.
}
\label{fig:ncoll_bj}
\end{figure}

In Fig.~\ref{fig:ncoll_bj},
we compare the $dN/dy$ dependence of the total number of collisions per $dN/dy$.
We consider three collision schemes: covariant,
CS-B, and CS-O (see more details in the previous section).
While these schemes exhibit similar behavior in the box calculation,
they show noticeable differences in the expanding system.
The covariant collision scheme predicts the same number of collisions for both
the collider frame and the target frame as it should,
while the number of collisions in the target frame is slightly smaller in the
non-covariant collision schemes, CS-B and CS-O\@.
This suppression of collision number in the target frame is also reported in
Ref.~\cite{Zhang:1996gb}.
The number of collisions in CS-B becomes very small
compared to the covariant scheme
in the expanding system, contrary to the static box calculations.
Collision scheme CS-O predicts a slightly larger collision number than the covariant scheme in the collider frame, although the difference between the
covariant scheme and CS-O is very small.
Let us consider a collision of two particles with $t_1$ and $t_2$
(assuming $t_1 < t_2$ without loss of generality)
where the times of the closest approach are given by $t_{c1}$ and $t_{c2}$.
In collision scheme CS-B, the collision time is assumed to be the minimum
of the times of the closest approach. As discussed in the previous section,
CS-B requires the condition $t_{c1} > t_2$
to prevent non-causal collisions.
The times of all particles are different in the initial condition~\eqref{eq:iniBJ}
unlike the box initial conditions, where the initial times of all particles are set to zero.
This expanding initial condition might be a part of the reasons
for the reduction of the collision number in CS-B\@.
A possible reason for the slight enhancement of the collision number in CS-O
may be due to the effects of the flow because CS-O computes the time of the closest
approach in the computational frame, 
while the others compute in the two-body c.m.s\@.

\subsection{AA and pA collisions}

In the preceding section, we examined different collision schemes for
a one-dimensionally expanding system assuming only elastic scatterings.
In this section,
we extend our investigation to include other effects
such as particle production, decay, and
the compression stages of nuclear collisions,
which involve more violent reactions compared to the expanding stages.
For this purpose,
we examine the central Au + Au (Pb + Pb) and p + Au collisions
using the event generator JAM~\cite{Nara:1999dz,Nara:2019crj},
in which particle productions are modeled by the excitations of resonances and
strings, followed by their decays.
The nuclear collisions are simulated
from the compression stages to the expanding stages, covering the entire
collision process until all the particles
freeze out.

\begin{figure}[hbt]
\includegraphics[width=9.0cm]{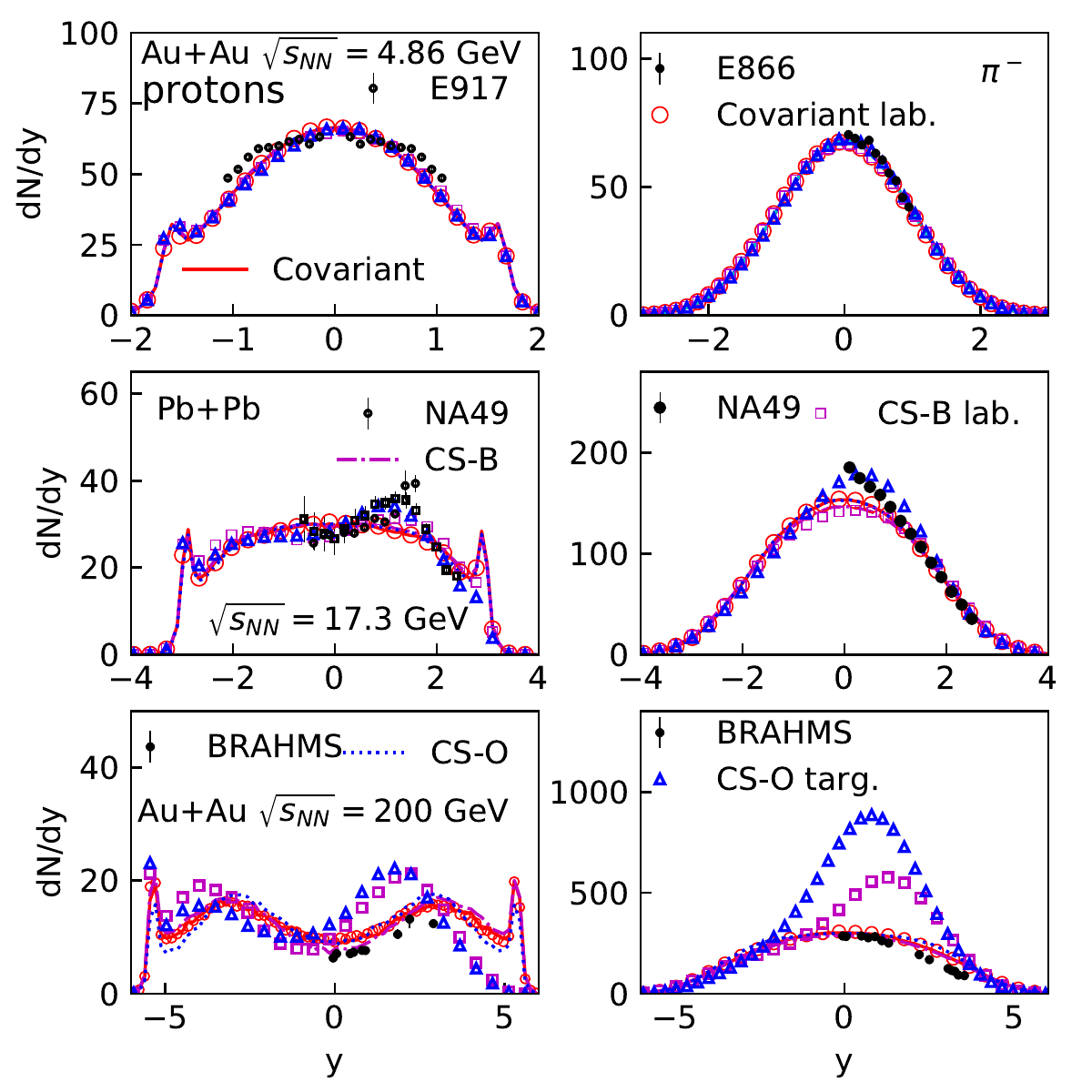}%
\comment{
\includegraphics[width=4.5cm]{dndy10_p.pdf}%
\includegraphics[width=4.5cm]{dndy10_pim.pdf}%

\includegraphics[width=4.5cm]{dndy158p.pdf}%
\includegraphics[width=4.5cm]{dndy158pim.pdf}%

\includegraphics[width=4.5cm]{dndy200_netp.pdf}%
\includegraphics[width=4.5cm]{dndy200_pim.pdf}%
}
\comment{
\includegraphics[width=5.5cm]{dndy10_p.pdf}%
\includegraphics[width=5.5cm]{dndy158p.pdf}%
\includegraphics[width=5.5cm]{dndy200_netp.pdf}%

\includegraphics[width=5.5cm]{dndy10_pim.pdf}%
\includegraphics[width=5.5cm]{dndy158pim.pdf}%
\includegraphics[width=5.5cm]{dndy200_pim.pdf}%
}
\caption{Rapidity distributions of protons and negative pions
in central Au + Au collisions
at $\sqrt{s_{NN}}=4.86$ GeV (upper panels)
$\sqrt{s_{NN}}=200$ GeV (lower  panels),
and
central Pb + Pb collision
at $\sqrt{s_{NN}}=17.3$ GeV (middle panels)
are compared with different collision schemes.
The solid, dashed-dotted, and dotted lines correspond to
the results of the calculations
from the covariant collision scheme, CS-B, and CS-O
in the center-of-mass frame.
The results of calculations in the target frame are
presented by circles, squares, and triangles
in the covariant collision scheme, CS-B, and CS-O, respectively.
Experimental data were taken from~\cite{Back:2002ic,E-802:1998xum,NA49:2010lhg,NA49:1998gaz,NA49:2002pzu,BRAHMS:2003wwg,BRAHMS:2004dwr}.
}
\label{fig:dndy10}
\end{figure}
\begin{figure}[htb]
\includegraphics[width=8.5cm]{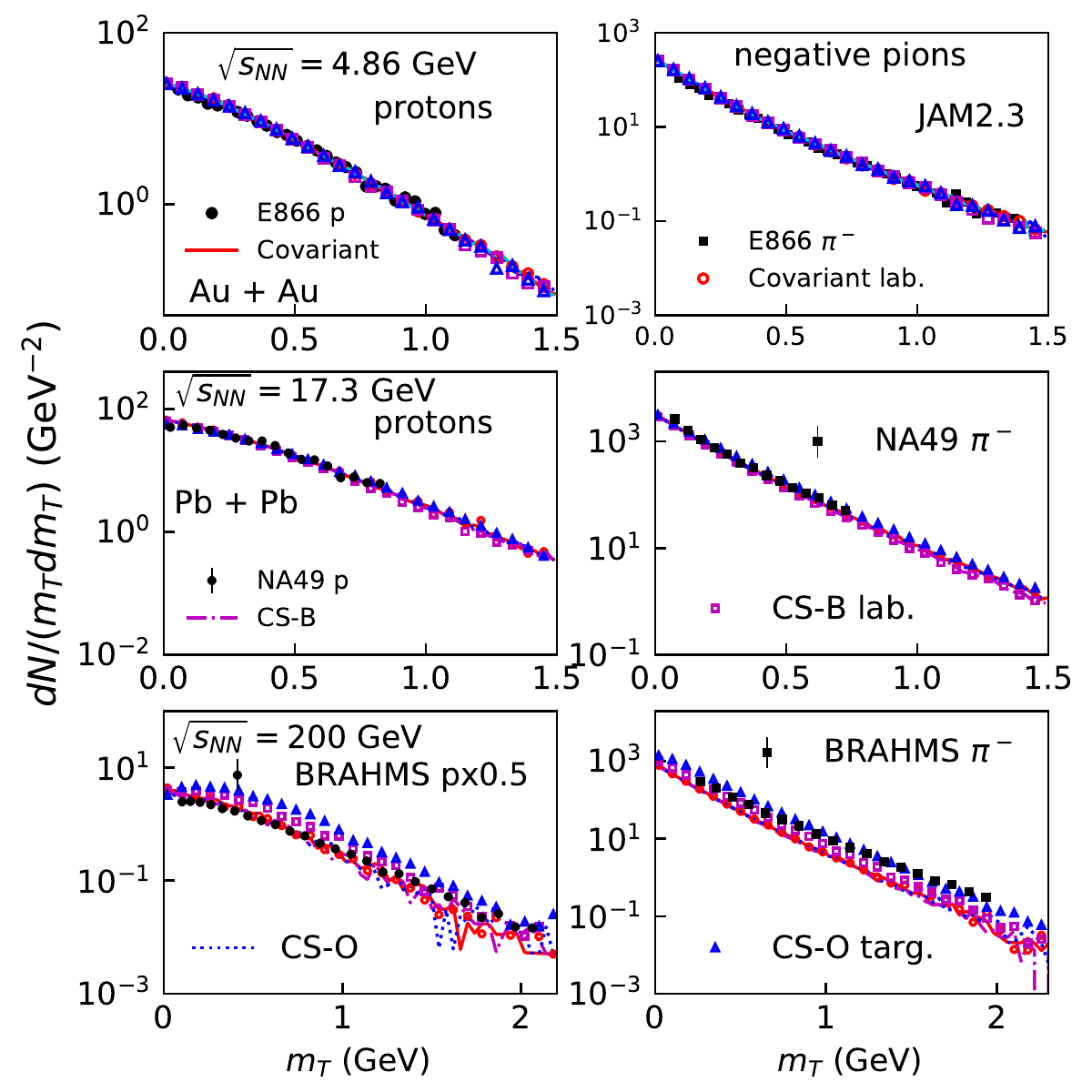}
\comment{
\includegraphics[width=4.5cm]{dndmt10_p.pdf}%
\includegraphics[width=4.5cm]{dndmt10_pim.pdf}%

\includegraphics[width=4.5cm]{dndmt158p.pdf}%
\includegraphics[width=4.5cm]{dndmt158pim.pdf}%

\includegraphics[width=4.5cm]{dndmt200p.pdf}%
\includegraphics[width=4.5cm]{dndmt200pim.pdf}%
}
\caption{
Transverse mass distributions of protons and negative pions
in central Au + Au collision at $\sqrt{s_{NN}}=4.86$ GeV (upper panels),
in central Pb + Pb collision at $\sqrt{s_{NN}}=17.3$ GeV (middle panels),
and in central Au + Au collision at $\sqrt{s_{NN}}=200$ GeV (lower panels).
The BRAHMS proton data are divided by factor 2 to correct weak
decay contribution roughly.
The meaning of the lines and symbols are the same as in Fig.~\ref{fig:dndy10}.
Experimental data were taken
from~\cite{E-802:1998xum,NA49:2006gaj,NA49:2002pzu,BRAHMS:2003wwg,BRAHMS:2004dwr}.
}
\label{fig:dndmt10}
\end{figure}

Figures~\ref{fig:dndy10} and~\ref{fig:dndmt10} show the
rapidity and transverse mass distributions of protons and negative pions
in central Au + Au collisions
at $\sqrt{s_{NN}}=4.86$ GeV (upper panels),
$\sqrt{s_{NN}}=200$ GeV (lower  panels),
and central Pb + Pb collisions
at $\sqrt{s_{NN}}=17.3$ GeV (middle panels)
calculated by the JAM2 cascade mode with different collision schemes.
The impact parameter $b\leq 3.4$ fm is chosen to approximately simulate the 5\%
central collisions.
The three different collision schemes, covariant, CS-B, and CS-O, are compared
including the frame dependence.
The three collision schemes yield almost identical rapidity and transverse mass distributions
for protons and pions for all three beam energies
when simulating in the global c.m.s\@.
However, a weak frame dependence is observed for the collision at
$\sqrt{s_{NN}}=4.86$ GeV\@.
The frame dependence becomes more pronounced at $\sqrt{s_{NN}}=17.3$~GeV
and even stronger at $\sqrt{s_{NN}}=200$~GeV\@.
The frame dependence of scheme CS-O turned out to be more significant
compared with scheme CS-B\@.


\begin{figure}[hbt]
\includegraphics[width=8.5cm]{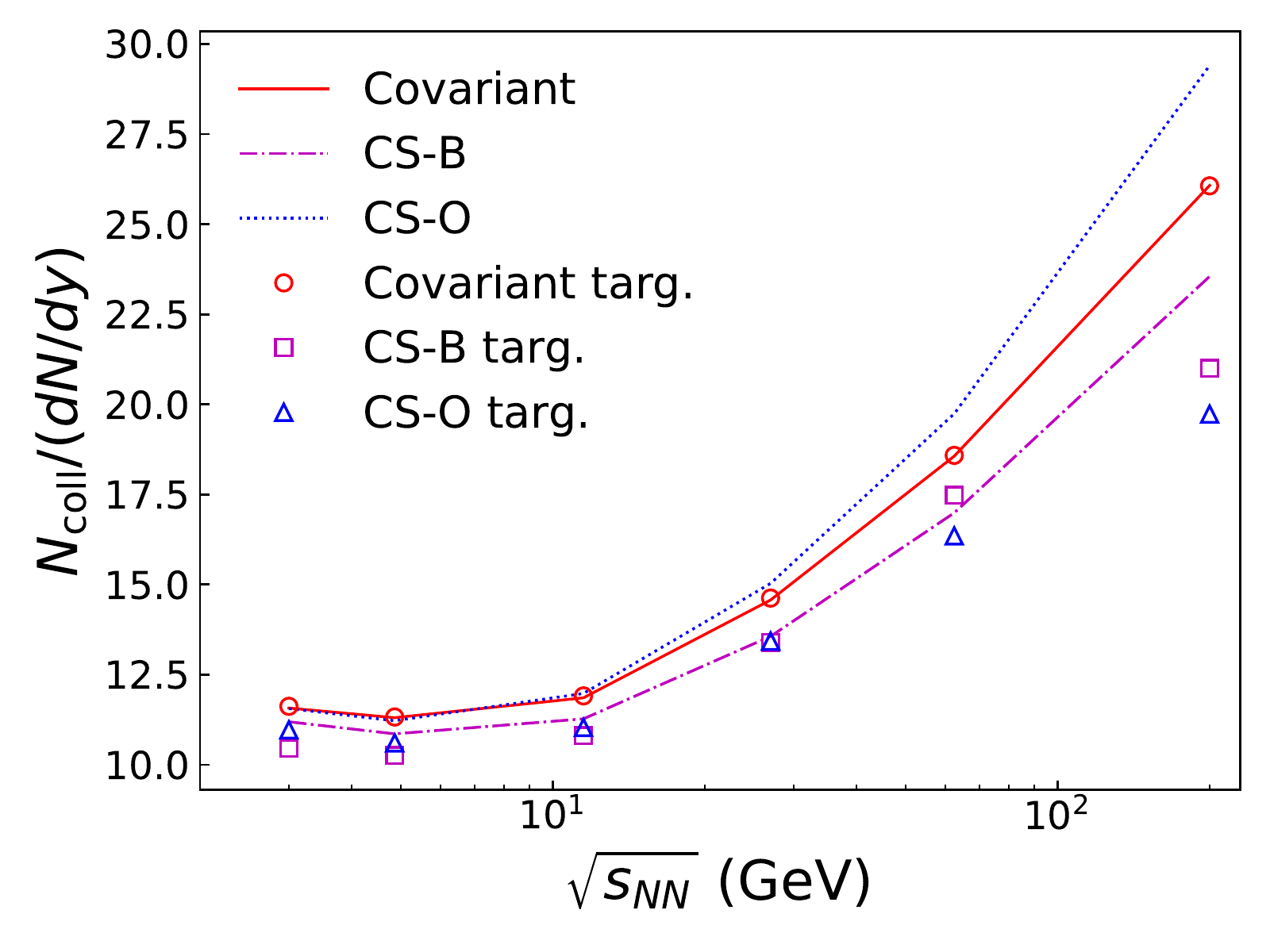}%
\caption{Collision energy dependence of the total number of collisions in central Au + Au collision ($b\leq 3.4$ fm)
normalized by the $dN/dy$ for all particles at $|y|<0.2$.
Symbols represent the results from the laboratory frame or the target frame,
while lines show the results of the simulation in the global c.m.s\@.
}
\label{fig:ncollAuAu}
\end{figure}

To see the difference in the collision schemes in more detail,
we display in Fig.~\ref{fig:ncollAuAu}
the beam energy dependence of the total number of collisions per $dN/dy$ at
mid-rapidity ($|y|<0.2$) for different collision schemes.
We observe that the collision number for scheme CS-O is nearly identical
to the covariant scheme up to the center-of-mass energy of $\sqrt{s_{NN}}=20$ GeV
when a simulation is done in the global c.m.s\@.
On the other hand, the collision number in scheme CS-B consistently appears
below the results obtained from the covariant scheme.
Furthermore, we observe that
the difference in the collision number between the c.m.s. and laboratory (target) frame is more pronounced in scheme CS-O compared with scheme CS-B, which
leads to the strong frame dependence seen in the particle spectra.

\begin{figure}[hbt]
\includegraphics[width=8.5cm]{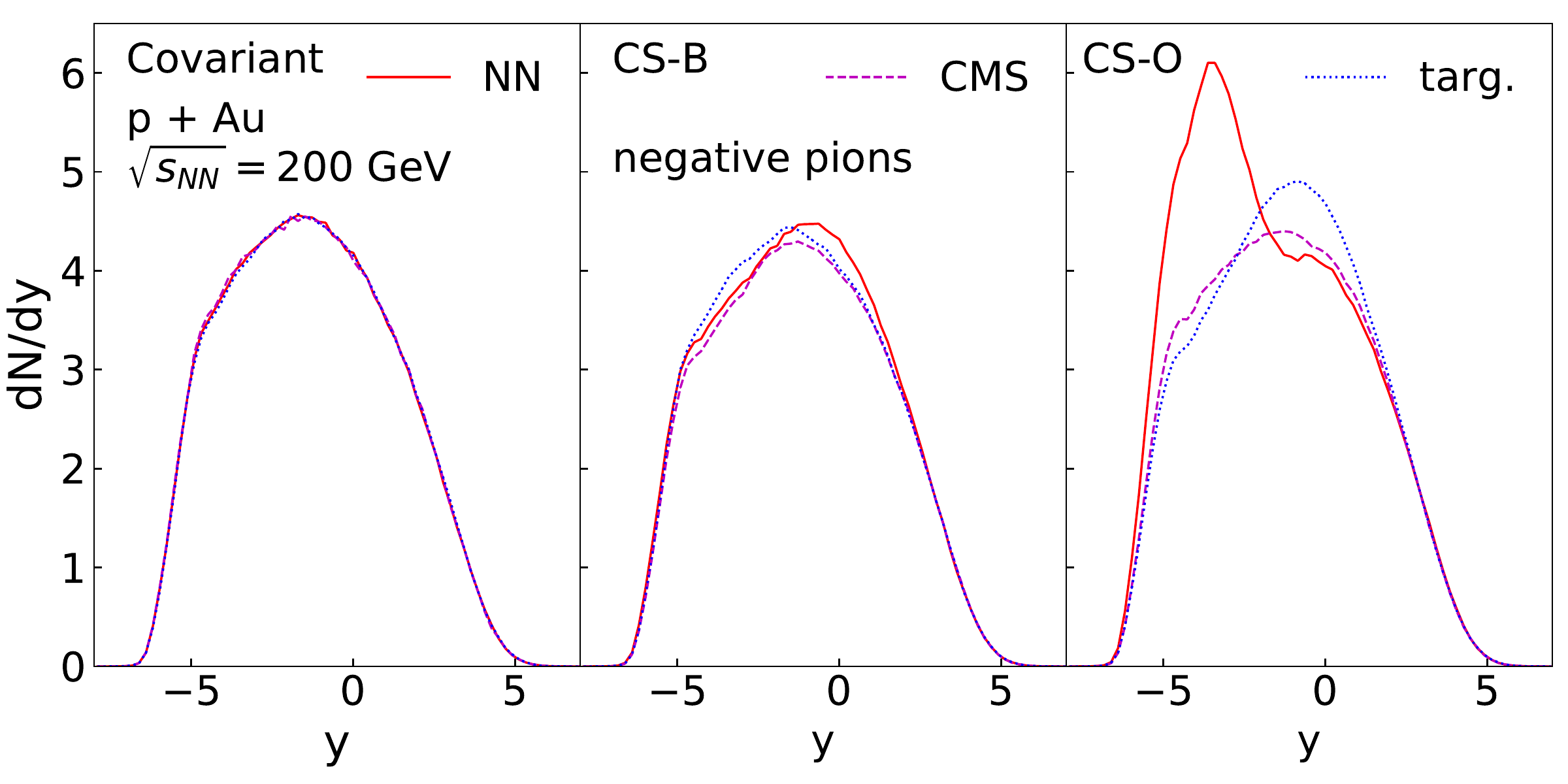}%
\caption{Rapidity dependence of negative pions in p + Au collisions ($b=1$~fm)
at $\sqrt{s_{NN}}=200$~GeV from the collision schemes Covariant (left panel),
CS-B (middle panel), and CS-O (right panel) are compared for
different frames. The equal-speed frame (NN frame),
the global c.m.s., and the target frame
are plotted by solid, dashed, and dotted lines, respectively.
}
\label{fig:pAu}
\end{figure}

We now examine asymmetric nuclear collisions.
Figure~\ref{fig:pAu} shows the rapidity distribution of negative pions
in central p + Au collisions at $\sqrt{s_{NN}}=200$ GeV\@.
The results from the three computational frames---the equal-speed frame of
the two colliding nuclei (NN), the global c.m.s.,
and the target frame---are compared for the collision schemes: covariant,
CS-B, and CS-O\@.
It is observed that the results from the three schemes are close to one
another in the global c.m.s\@.
The scheme CS-O predictions show strong frame dependence
while scheme CS-B does not.
The study of the frame dependence in p + Au collisions
reveals that non-covariant cascade schemes perform best in the global c.m.s.
but not in the equal-speed frame in pA collisions.

\subsection{Cluster separability}
\label{sec:cluster}

In this section, we address the issue of the cluster separability
in our covariant cascade scheme.
When a system is separated into independent clusters, clusters do not interact with
each other in our approach since the timekeeper is a constant vector.
However,
our scheme does not guarantee the cluster separability
in the following sense:
When the system is divided into two subsystems with the total momentum
of $P_1$ and $P_2$,
\begin{equation}
 P = P_1 + P_2,
\end{equation}
the cluster separability condition implies that
the equations of motion for the system
must exhibit two distinct sets of equations,
\begin{align}
\frac{dq_i}{d\tau} &= \frac{p_i}{\hat{a}_1\cdot p_i}, &
\frac{dq_j}{d\tau} &= \frac{p_j}{\hat{a}_2\cdot p_j},
\label{eq:cluster_sep}
\end{align}
where $\hat{a}_1=P_1/\sqrt{P_1^2}$ and $\hat{a}_2=P_2/\sqrt{P_2^2}$.
However, this condition is not satisfied in our model.
To investigate the effects of cluster separability,
we manually impose the cluster separability condition
in numerical simulations:
when clusters are identified, we incorporate Eq.~\eqref{eq:cluster_sep}
into the actual simulations.
To identify clusters we introduce time steps,
and at every time step, we group particles that are close
to one another in the coordinate space based on the interaction range.

\begin{figure}[tbh]
\includegraphics[width=8.5cm]{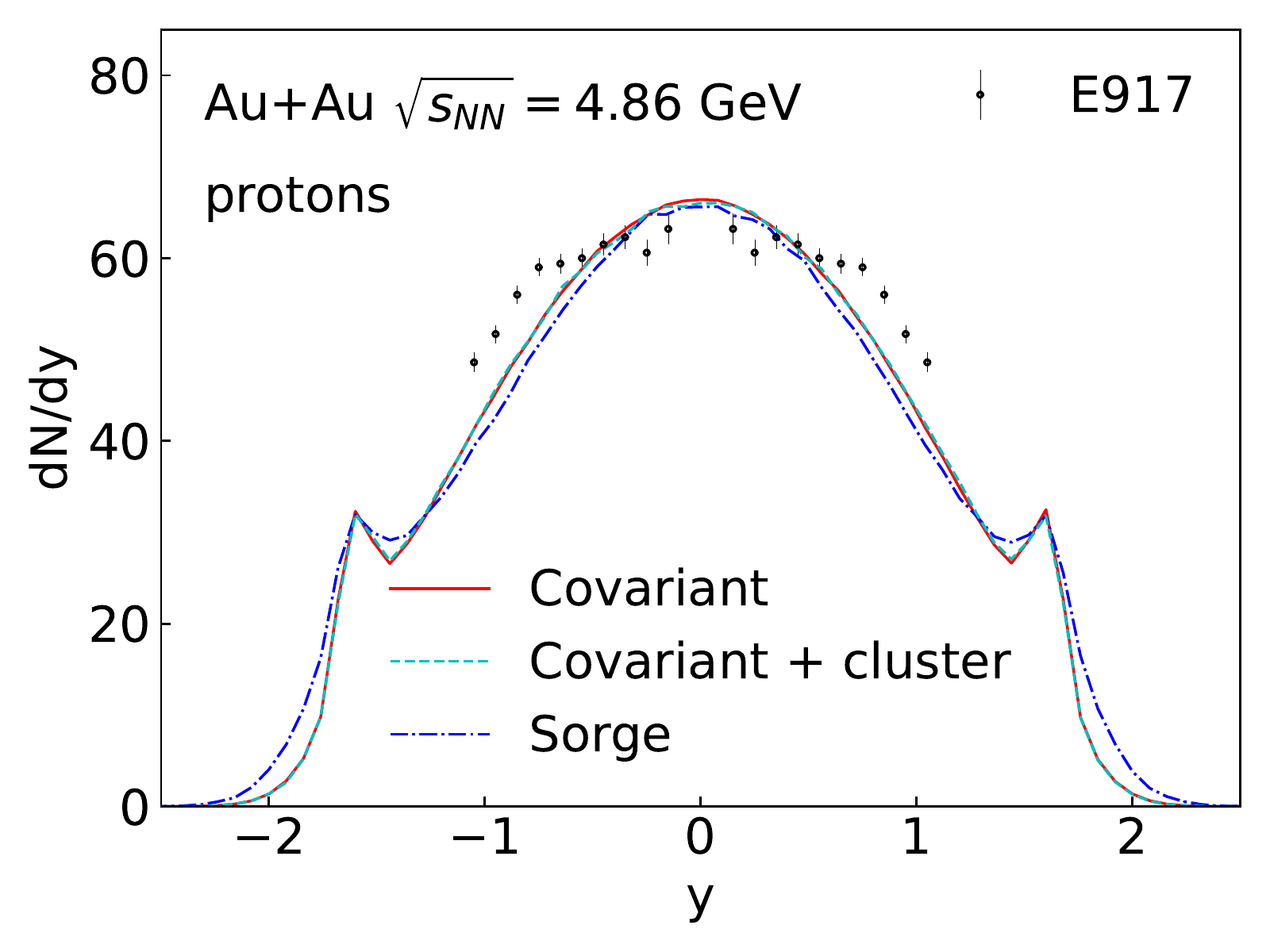}%
\caption{Rapidity distributions in central Au + Au collision at
$\sqrt{s_{NN}}=4.86$ GeV are compared with different covariant collision schemes.
The solid and dashed lines show the results from the covariant method
without and with cluster separation, respectively.
The dashed-dotted line corresponds to the results from the covariant method
using Sorge's constraints~\eqref{eq:SorgeConst}.
Experimental data were taken from~\cite{Back:2002ic}.
}
\label{fig:dndy120}
\end{figure}

The rapidity distribution of protons from our approach
is compared with the original covariant method in Fig.~\ref{fig:dndy120}.
We found that our original equations of motion yield the
same results as the one that the cluster separability is imposed.
As a comparison, the rapidity distribution from the model with
Sorge's constraints~\eqref{eq:SorgeConst} is plotted by a dashed-dotted line in
Fig.~\ref{fig:dndy120}, which is in good agreement with our covariant method.
Based on these findings, it appears that the issues of cluster separability may not be highly relevant
for simulating high-energy nuclear collisions using our approach.
However, it is worth noting that when nuclear cluster productions are discussed with
the model, such as in the relativistic quantum molecular dynamics,
the model incorporating cluster separability may produce more stable clusters.
We will leave this topic as a future work.

\section{Potential interactions}
\label{sec:potential}


The main ingredients of the QMD model are the Boltzmann-type collision term
to simulate hard interactions and the potential interactions for the soft part
of the interactions, which play a significant role in determining the collective
dynamics of the system.
The relativistic version of the QMD (RQMD) model has been developed
based on the relativistic constrained dynamics~\cite{Sorge:1989dy,Maruyama:1991bp}
using the constraints~\eqref{eq:SorgeConst}.
However, as we mentioned before, accurately solving the constraints~\eqref{eq:SorgeConst}
is practically infeasible with the current computer capabilities.
To address this challenge,
RQMD models with the time constraints~\eqref{eq:timefix}
were proposed~\cite{Maruyama:1996rn,Mancusi:2009zz,Nara:2019qfd,Nara:2020ztb,Nara:2021fuu},
which provide a numerically efficient method.
However, these models are limited to the uses in the global c.m.s\@.
We derive covariant equations of motion for the RQMD approach, which allows
numerically efficient simulations.

We consider the following mass shell constraints for the particle system
interacting through the Lorentz scalar $S_i$ and vector $V_i$ potentials,
\begin{equation}
 H_i= p_i^{*2} - m_i^{*2}, \quad i=1,\ldots, N,
\end{equation}
where $p_i^{*}=p_i - V_i$ and $m_i^* = m_i - S_i$.
We use the time constraints of Eqs.~\eqref{eq:timefix} and~\eqref{eq:timefix2}.
Under the assumption of the approximate commutation of the mass shell constraints
$[H_i,H_j]=0$ as in Ref.~\cite{Sorge:1989dy},
the equations of motions~\eqref{eq:EOM1} and~\eqref{eq:EOM2} become
\begin{align}
  \frac{dq^\mu_i}{d\tau} &= \sum_{j=1}^N 2u_j\Pi_{ji}^\mu, &
  \frac{dp^\mu_i}{d\tau} &= \sum_{j=1}^N 2u_j Q_{ji}^\mu,
\end{align}
where
\begin{align}
 \Pi_{ji}^\mu &= \frac{1}{2}\frac{\partial H_j}{\partial p_{i\mu}}
  = p_i^{*\mu}\delta_{ij}
     - {m_j^*}
      \frac{\partial m_j^*}{\partial p_{i\mu}}
      -p^{*\nu}_j
      \frac{\partial {V}_{j\nu}}{\partial p_{i\mu}}, \\
 Q_{ji}^\mu &=
-\frac{1}{2}\frac{\partial H_j}{\partial q_{i\mu}}
      =
      m_j^*
      \frac{\partial m_j^*}{\partial q_{i\mu}}
      +p^{*\nu}_j
     \frac{\partial V_{j\nu}}{\partial q_{i\mu}} .
\end{align}
We need to solve the following system of equations for the Lagrangian multipliers
$u_i$:
\begin{equation}
 \sum_{j=1}^N  (\hat{a} \cdot
  \Pi_{ji})\,
  u_j =
 1, \quad i=1,\ldots,N\,.
\label{eq:umatrix}
\end{equation}

Let us consider some approximations to avoid solving the system of equations.
If we neglect the derivatives of potentials with respect to the energy,
we have only the free part,
$\hat{a}_i\cdot(\partial H_j/\partial p_i)=2p_i^{*0}$ in the frame
specified by $\hat{a}=(1,0,0,0)$.
Thus, the Lagrange multipliers $u_i$ are given by
$u_i=1/(2\hat{a}\cdot p_i^*)$ in any computational frame,
and we get the equations of motion
\begin{align}
  \frac{dq^\mu_i}{d\tau} &= \sum_{j=1}^N \frac{\Pi_{ji}^\mu}{\hat{a}\cdot p_j^*}, &
  \frac{dp^\mu_i}{d\tau} &= \sum_{j=1}^N \frac{Q_{ji}^\mu}{\hat{a}\cdot p_j^*}.
\end{align}
These equations of motion with $\hat{a}=(1,0,0,0)$ are successfully
applied in Refs.~\cite{Nara:2019qfd, Nara:2020ztb, Nara:2021fuu}
for the simulations of heavy-ion collisions.

We expect that the off-diagonal parts of Eq.~\eqref{eq:umatrix} are small because only one term of the derivatives
in the sum of the potentials is non-zero, unlike the diagonal term, in which we
need to take the sum for all particles for the derivatives.
When we only take the diagonal parts,
$u_i$ can be obtained trivially as $u_i=1/(2\hat{a}\cdot \Pi_{ii})$,
and the equations of motion become
\begin{align}
\frac{dq^\mu_i}{d\tau} &= \sum_{j=1}^N
\frac{\Pi^\mu_{ji}}
    {\hat{a}\cdot \Pi_{jj}}, &
\frac{dp^\mu_i}{d\tau}=
  \sum_{j=1}^N \frac{Q_{ji}^\mu}{\hat{a}\cdot \Pi_{jj}}.
\label{eq:rqmd2}
\end{align}
We note that the diagonal part $\Pi_{ii}$
is the same as $\tilde{\Pi}_\mu$ in Eq.~(33) in Ref.~\cite{Weber:1992qc}.
When we take only $j=i$ part of the sum in Eq.~\eqref{eq:rqmd2} in the frame $\hat{a}=(1,0,0,0)$,
where all time coordinates of particles are the same as $\tau$,
Eq.~\eqref{eq:rqmd2} becomes identical to the equations of motion of the test particles
in the relativistic BUU (RBUU) approach~\cite{Weber:1992qc}.
The numerical simulations of the RBUU approach with these equations of motion have been
realized for the study of heavy-ion collisions in Ref.~\cite{Maruyama:1993jb}.

\section{Summary}
\label{sec:summary}

We have presented a Poincar\'e covariant cascade method that enables
efficient numerical simulation of the Boltzmann-type collision term.
This method provides an effective approach for accurately modeling and studying collision processes
in high energy heavy-ion collisions.
We have verified that our covariant cascade method predicts the correct collision rate
and the thermal spectrum in a box simulation.
Moreover, we have demonstrated the frame independence of our method in a one-dimensionally expanding system
as well as actual nuclear collisions, including AA and pA collisions.

Detailed comparisons with the cascade schemes in the $6N$-dimensional phase space have been conducted.
It is found that some non-covariant collision schemes can yield
reliable results under specific conditions.
Specifically, collision scheme CS-O shows reliable results
when applied to an expanding system
or to the collision of two nuclei with a beam energy less than $\sqrt{s_{NN}} \simeq 20$ GeV\@.
In CS-O, the impact parameter is defined in the two-body c.m.s. of the colliding particles,
and the collision time is specified
in the computational frame, which is chosen to be the global center-of-mass frame
of the two nuclei.

Finally, we proposed numerically efficient covariant equations of motion for $N$-particle systems
interacting through potentials,
which can be utilized in the QMD simulations at relativistic collision energies.

\begin{acknowledgments}
This work was supported in part by the
Grants-in-Aid for Scientific Research from JSPS
(Nos. JP21K03577, 
JP19H01898, 
JP21H00121, 
and JP23K13102).
This work was also supported by JST,
the establishment of university fellowships towards
the creation of science technology innovation,
Grant Number JPMJFS2123.
\end{acknowledgments}

\appendix

\section{Closest distance approach in $6N$-dimensional phase space}
\label{appendix:CD}

In this appendix, we provide a derivation of the expressions
for the impact parameter and the times of the closest
approach within a $6N$-dimensional phase space approach.
See also Refs.~\cite{Cheng:2001dz,Sasaki:2000pu,Hirano:2012yy}.

The impact parameter is defined as
the minimum distance in the two-body center-of-mass system (c.m.s.),
\begin{equation}
  b^2 = \bm{x}_{\cm}^2 - \frac{(\bm{x}_{\cm}\cdot\bm{v}_{\cm})^2}{\bm{v}_{\cm}^2}.
\label{eq:bcm}
\end{equation}
To obtain the Lorentz invariant expression, we define
the transverse relative distance $x_T$ and the transverse relative momentum
$p_T$:
\begin{align}
   q_T &\equiv \Delta^{\mu\nu} q_\nu =  q - (q\cdot u)u,\\
   p_T &\equiv \Delta^{\mu\nu} p_\nu =  p - (p\cdot u)u,
\end{align}
where
$q= q_1-q_2$, $p=p_1 - p_2$,
$u=\frac{p_1+p_2}{\sqrt{(p_1+p_2)^2}}$,
and $\Delta^{\mu\nu}=g^{\mu\nu}-u^\mu u^\nu$ is the projector
to the transverse distances.
The impact parameter vector $b$ is orthogonal to the transverse relative momentum
$b\cdot p_T=0$, which is given by
\begin{equation}
   b =  q_T - \frac{(q_T\cdot p_T)}{p_T^2}p_T
     =  q_T - \frac{(q\cdot p_T)}{p_T^2}p_T,
\label{eq:impactparametervector}
\end{equation}
where we used $q_T\cdot p_T = q\cdot p_T$.
Then, the square of the invariant impact parameter is evaluated as
~\cite{Kodama:1983yk,Sasaki:2000pu,Hirano:2012yy}
\begin{equation}
   -b\cdot b = -q_T^2 + \frac{(q\cdot p_T)^2}{p_T^2}
   = -q^2 +(q\cdot u)^2 + \frac{(q\cdot p_T)^2}{p_T^2}.
\end{equation}
Noting that the relative momentum is proportional to the relative velocity
in the c.m.s.,
this expression is equivalent to Eq.~\eqref{eq:bcm} since
\begin{align}
  \bm{x}_{\cm}^2 &= t_{\cm}^2 - q^2 = (q\cdot u)^2 - q^2,\\
  \bm{p}_{\cm}^2 &= E_{\cm}^2 - p^2 = (p\cdot u)^2 - p^2,\\
  \bm{x}_{\cm}\cdot\bm{p}_{\cm} &= E_{\cm}t_{\cm} - q\cdot p
   = q \cdot [(p\cdot u)u - p].
\end{align}

The times of the closest approach
are also computed in the two-body c.m.s.:
\begin{align}
t_{\cm} - t_{\cm,1} &=
-\frac{[\bm{x}_{\cm,1}(t_{\cm,1})-\bm{x}_{\cm,2}(t_{\cm,1})]\cdot\bm{v}_{\cm}}{\bm{v}_{\cm}^2}\nonumber\\
&= -\frac{[\bm{x}_{\cm}-\bm{v}_{\cm,2}t_{\cm}]\cdot\bm{v}_{\cm}}{\bm{v}_{\cm}^2},
\end{align}
where $t_{\cm}=t_{\cm,1}-t_{\cm,2}$.
Using the relation, 
\begin{equation}
\bm{v}_{\cm}
= -\bm{p}_{\cm,2}\left(\frac{1}{E_{\cm,1}}+\frac{1}{E_{\cm,2}}\right)
\end{equation}
and
\begin{equation}
(\bm{x}_{\cm}-\bm{v}_{\cm,2}t_{\cm})\cdot\bm{p}_{\cm,2}
= x\cdot p_2 - \frac{t_{\cm}}{E_{\cm,2}}p_2^2,
\end{equation}
a Lorentz invariant expression of the collision time is obtained as
\begin{align}
t_{\cm} - t_{\cm,1}
&=\frac{E_{\cm,1}}{E_{\cm}}\frac{t_{\cm}p_2^2 - E_{\cm,2}(x\cdot p_2) }{(p_2\cdot u)^2-p_2^2} \nonumber \\
&=E_{\cm,1}\frac{(x\cdot p_1)p_2^2 - (x\cdot p_2)(p_1\cdot p_2) }
               {(p_1\cdot p_2)^2-p_1^2p_2^2}
\end{align}
where $E_{\cm}=E_{\cm,1}+E_{\cm,2}=\sqrt{(p_1+p_2)^2}$.
The times of the closest approach $t_{c1}$ and $t_{c2}$ for the two colliding particles
in the computational frame
are obtained by Lorentz transforming the collision time
\begin{equation}
  t_{c1} - t_1 = (t_{\cm}-t_{\cm,1})\frac{E_1}{E_{\cm,1}},
\end{equation}
which are given by
\begin{align}
 t_{c1} &= t_1 +\frac{p_2^2(q\cdot p_1) - (p_1\cdot p_2)(q\cdot p_2)}
                         {(p_1p_2)^2 - p_1^2p_2^2} E_1,\\
 t_{c2} &= t_2 -\frac{p_1^2(q\cdot p_2) - (p_1\cdot p_2)(q\cdot p_1)}
                            {(p_1p_2)^2 - p_1^2p_2^2} E_2.
\end{align}

\section{Impact parameter}
\label{appendix:IP}

In this appendix,
we show that
the impact parameter Eq.~\eqref{eq:impactpara} is equivalent to
the transverse distance at the collision point in the Lorentz covariant
cascade method in  Eq.~\eqref{eq:impact}.
From the equations of motion for two particles Eq.~\eqref{eq:ceom} and
the time of closest approach~Eq.~\eqref{eq:ctime1},
the collision point is expressed as
\begin{equation}
 q_c = 
q' -v_2'(\tau_1-\tau_2),
\end{equation}
where we have used the relation $v\cdot v_T = v_T^2$, and
\begin{align}
 q' &= q - \frac{(q\cdot v_T)}{v_T^2}v, &
 v'_2 &= v_2 - \frac{(v_2\cdot v_T)}{v_T^2}v.
\end{align}
The impact parameter vector becomes
\begin{align}
 b &= q_c - \frac{(q_c\cdot P)}{P^2}P \nonumber \\
   &= q_T - \frac{(q_T\cdot v_T)}{v_T^2}v_T
  - v_{2T}'\,(\tau_1 - \tau_2)
\end{align}
where
\begin{align}
  v_{T2} &= v_2 - \frac{(v_2\cdot P)}{P^2}P, &
  v_{T2}' &= v_{T2} - \frac{(v_{T2}\cdot v_T)}{v_T^2}v_T.
\label{eq:vt2d}
\end{align}
Noting that the transverse relative velocity $v_{T}=v_{T1}-v_{T2}$
is parallel to the transverse velocity of a particle 2 $v_{T2}$,
\begin{equation}
  v_T \xrightarrow[\text{c.m.}]{} \bm{v}_{\cm}=\bm{v}_{\cm,1}-\bm{v}_{\cm,2}
  \approx -\bm{v}_{\cm,2},
\end{equation}
we can replace
$v_{T}$ by $v_{2T}$ in Eq.~\eqref{eq:vt2d}, and
one finds $v'_{T2}=0$.
Furthermore, since $v_T$ is also parallel to $p_T$,
the impact parameter vector becomes
\begin{equation}
 b =  q_T - \frac{(q\cdot v_T)}{v_T^2}v_T
   =  q_T - \frac{(q\cdot p_T)}{p_T^2}p_T,
\end{equation}
which is the same expression as the impact parameter vector in the c.m.s.
Eq.~\eqref{eq:impactparametervector}.

\bibliography{ref}

\end{document}